\documentclass[12pt]{article}
\usepackage{amssymb,amsmath} 
\usepackage{graphicx}
\usepackage{epsfig} 
\usepackage{latexsym}
\usepackage{psfrag}   
\usepackage{subfigure}  
\usepackage{booktabs}  
\usepackage{braket}
\usepackage{mathtools}
\usepackage{textcomp}
\usepackage{protosem}
\usepackage{wasysym}
\usepackage[numbers,sort&compress]{natbib}
\usepackage{mathrsfs}
\usepackage[inner=2.9cm,outer=2cm]{geometry}
\usepackage[toc]{appendix}
\usepackage{color,soul} 
\usepackage{datetime}
\usepackage[
      colorlinks=true,
      linkcolor=darkblue,  
      urlcolor=black,    
      filecolor=blue,     
      citecolor=darkgreen, 
      linktocpage=true,
      pdfstartview=FitV,
      bookmarksopen=true    
      ]{hyperref}

\definecolor{darkblue}{rgb}{0.2, 0, 0.8}
\definecolor{darkgreen}{rgb}{0.2, 0.71, 0}

\numberwithin{equation}{section}

\newcommand{\req}[1]{(\ref{#1})} 
\newcommand{\labell}[1]{\label{#1}}

\newcommand{\bea}{\begin{eqnarray}}
\newcommand{\eea}{\end{eqnarray}}
\newcommand{\ba}{\begin{eqnarray}}
\newcommand{\ea}{\end{eqnarray}}

\newcommand{\beq}{\begin{equation}}
\newcommand{\eeq}{\end{equation} }
\newcommand{\beqa}{\begin{eqnarray}}
\newcommand{\eeqa}{\end{eqnarray}}
\newcommand{\beqar}{\begin{eqnarray*}}
\newcommand{\eeqar}{\end{eqnarray*}}

\newcommand{\reef}[1]{(\ref{#1})}
\newcommand{\ssc}{\scriptscriptstyle}
\newcommand{\eg}{{\it e.g.,}\ }
\newcommand{\ie}{{\it i.e.,}\ }

\newcommand{\E}{\mathcal{E}}

\newcommand{\ctt}{C_{\ssc T}}

\DeclareMathOperator{\tr}{Tr}  

\renewcommand{\href}[2]{#2}

\newenvironment{changemargin}[2]{%
\begin{list}{}{%
\setlength{\topsep}{0pt}%
\setlength{\leftmargin}{#1}%
\setlength{\rightmargin}{#2}%
\setlength{\listparindent}{\parindent}%
\setlength{\itemindent}{\parindent}%
\setlength{\parsep}{\parskip}%
}%
\item[]}{\end{list}}

\begin{document}  


\begin{titlepage}



\vspace*{1.2cm}

\begin{center}
{\LARGE \bf f(Lovelock) theories of gravity} \\

\vspace*{1.2cm}
{\bf Pablo Bueno$^{\text{\lightning}}$, Pablo A. Cano$^{ \textproto{\AAdaleth}}$, \'Oscar Lasso A.$^{\textproto{\AAdaleth}}$ and Pedro F. Ram\'irez$^{\textproto{\AAdaleth},\text{\eighthnote}}$}
\medskip\vspace{0.5cm}
$^{\text{\lightning}}$Instituut voor Theoretische Fysica, KU Leuven\\ 
Celestijnenlaan 200D, B-3001 Leuven, Belgium\bigskip

$^{\textproto{\textproto{\AAdaleth}}}$Instituto de F\'isica Te\'orica UAM/CSIC \\
C/ Nicol\'as Cabrera, 13-15, C.U. Cantoblanco, 28049 Madrid, Spain\bigskip

$^{\text{\eighthnote}}$Institut de Physique Th\'eorique, Universit\'e Paris Saclay, CEA\\ CNRS, F-91191 Gif-sur-Yvette, France\\

\bigskip

\end{center}

\vspace*{0.1cm}

\begin{abstract}  
\begin{changemargin}{-0.95cm}{-0.95cm}
$f($Lovelock$)$ gravities are simple generalizations of the usual $f(R)$ and Lovelock theories in which the gravitational action depends on some arbitrary function of the corresponding dimensionally-extended Euler densities. In this paper we study several aspects of these theories in general dimensions. We start by identifying the generalized boundary term which makes the gravitational variational problem well-posed. Then, we show that these theories are equivalent to certain scalar-tensor theories and how this relation is characterized by the Hessian of $f$. We also study the linearized equations of the theory on general maximally symmetric backgrounds. Remarkably, we find that these theories do not propagate the usual ghost-like massive gravitons characteristic of higher-derivative gravities on such backgrounds. In some non-trivial cases, the additional scalar associated to the trace of the metric perturbation is also absent, being the usual graviton the only dynamical field. In those cases, the linearized equations are exactly the same as in Einstein gravity up to an overall factor, making them appealing as holographic toy models. We also find constraints on the couplings of a broad family of five-dimensional $f($Lovelock$)$ theories using holographic entanglement entropy. Finally, we construct new analytic asymptotically flat and AdS/dS black hole solutions for some classes of $f($Lovelock$)$ gravities in various dimensions.
\end{changemargin}
\end{abstract} 

\end{titlepage}

\setcounter{tocdepth}{2}
{\small
\setlength\parskip{-0.5mm} 
\noindent\rule{15.7cm}{0.4pt}
\tableofcontents
\vspace{0.6cm}
\noindent\rule{15.7cm}{0.4pt}
}

\section{Introduction \& summary of results} 
\label{sec:Introduction} 
Higher-derivative theories of gravity have been subject of intense study in recent years. The reasons for this interest are diverse.
From a fundamental perspective, it is clear now that general relativity is an effective description which most certainly needs to be completed in the ultraviolet. A characteristic manifestation of the putative underlying theory would be the appearance of a series of higher-derivative terms, consisting of different contractions of the Riemann tensor and its covariant derivatives, which would correct the Einstein-Hilbert (EH) action at sufficiently high energies.
This is of course the case of String Theory, which generically predicts an infinite series of such terms --- see \eg \cite{Gross:1986mw,Green:2003an,Frolov:2001xr}.

While many explicit String Theory models giving rise to particular effective higher-derivative theories have been constructed, there exists a less fundamental but more practical approach which has been also vastly studied in the literature. Such approach consists in regarding certain higher-derivative theories as quantum gravity toy models. This is the case, for example, of \emph{topologically massive gravity} \cite{Deser:1981wh} and \emph{new massive gravity} \cite{Bergshoeff:2009hq} in three dimensions, or \emph{critical gravity} \cite{Lu:2011zk} in four. In these theories, and others of the like, the EH action is supplemented by a few additional higher-derivative terms which improve some of the properties of the original theory --- \eg by making it renormalizable \cite{Stelle:1977ry,Stelle:1976gc}.

Constructions of this kind are also very useful in the holographic context \cite{Maldacena,Gubser,Witten}. Indeed, through the holographic dictionary, higher-derivative theories have been successfully used to unveil various properties of general strongly coupled systems in various dimensions --- see \eg \cite{Myers:2010xs,Myers:2010tj,Brigante:2007nu,Bueno1,Bueno2,Mezei:2014zla}. In this context, the philosophy also consists in regarding these theories as computationally useful toy models: if a certain property holds for general strongly coupled conformal field theories (CFTs), it is reasonable to expect that these toy models are able to capture it --- and that has been proven to be very often the case. A paradigmatic example of this class of theories is \emph{quasi-topological gravity} \cite{Oliva:2010eb,Myers:2010ru}, which was precisely conceived as a multi-parameter holographic toy model of strongly coupled CFTs in various dimensions.

Most likely, the area of research in which higher-derivative gravities have appeared more often is cosmology. In that context, these terms are considered with the idea that general relativity might not be, after all, the right description of the gravitational interaction at cosmological scales. This is of course motivated by the puzzling existence of dark matter and dark energy, as well as by the need to construct a coherent picture --- beyond the $\Lambda$-CDM model --- of the universe evolution able to incorporate, in particular, an inflationary scenario compatible with the observations.\footnote{The body of literature in this area is huge. See \eg \cite{Nojiri:2006ri,Sotiriou:2008rp,Nojiri:2010wj,Clifton:2011jh} for some nice reviews on higher-derivative gravities and cosmology.}

Two of the higher-derivative theories which have received more attention within the areas explained above are Lovelock \cite{Lovelock1,Lovelock2} and $f(R)$ gravities --- see \eg \cite{Sotiriou:2008rp}. While there has been a large amount of papers studying different aspects of these higher-derivative gravities, remarkably little work has been done on the class of theories which most naturally incorporates both $f(R)$ and Lovelock in a common framework. We are talking, of course, about $f($Lovelock$)$ gravities, which are the subject of this paper.

The most general $f($Lovelock$)$ action can be written as
\begin{equation}\labell{flov}
S_{ f(\rm{Lovelock})}=\frac{1}{16\pi G}\int_{\mathcal{M}}d^Dx\,\sqrt{|g|}\,f(\mathcal{L}_0,\mathcal{L}_1,\dots ,\mathcal{L}_{\lfloor D/2 \rfloor})\, ,
\end{equation}
where $f$ is some differentiable function of the dimensionally extended Euler densities (ED)\footnote{The alternate Kronecker symbol is defined as: $\delta^{\mu_1\mu_2\dots \mu_r}_{\nu_1\nu_2\dots\nu_r}= r!\delta^{[\mu_1}_{\nu_1}\delta^{\mu_2}_{\nu_2}\dots \delta^{\mu_r]}_{\nu_r}$.
}
\begin{equation}
\mathcal{L}_{j}= \frac{1}{2^{j}}\delta^{\mu_1\dots \mu_{2j}}_{\nu_1\dots \nu_{2j}}R^{\nu_1\nu_2}_{\mu_1\mu_2}\dots R^{\nu_{2j-1}\nu_{2j}}_{\mu_{2j-1}\mu_{2j}}\, .
\end{equation}
 In particular, \eg $\mathcal{L}_1=R$,  and $\mathcal{L}_2=R^2-4R_{\mu\nu}R^{\mu\nu}+R_{\mu\nu\rho\sigma}R^{\mu\nu\rho\sigma}$, which are the usual EH and Gauss-Bonnet (GB) terms respectively. Note that $\mathcal{L}_j$ vanishes identically for $j> \lfloor D/2 \rfloor$, \ie  for $j>D/2$ when $D$ is even, and for $j>(D-1)/2$ for odd $D$.

Naturally, the above action \reef{flov} reduces to the usual Lovelock and $f(R)$ theories when we choose $f$ to be a linear combination of ED and some arbitrary function of the Ricci scalar respectively, 
\begin{equation}\label{cases}
f_{\text{Love.}}=\sum_{n=0}^{\lfloor D/2 \rfloor}\lambda_{n}\,\Lambda_0^{1-n}\, \mathcal{L}_n\, , \quad f_{f(R)}=f(R)\, ,
\end{equation}
where $\Lambda_0^{-1/2}$ is some length scale, and $\lambda_j$ are dimensionless couplings.\footnote{Note that explicit cosmological constant and EH terms can be trivially made appear in $f_{\text{Love.}}$ by setting: $\lambda_0= - 2$ and $\lambda_1=1$ respectively. Similarly, we could replace $f_{f(R)}$ in \req{cases} by $f_{f(R)}=-2\Lambda_0+R+f(R)$ to make those terms explicit in the $f(R)$ action.} When $D$ is even, the combination `$\sqrt{|g|}\,\mathcal{L}_{D/2}$' is topological in the sense that its integral over a boundaryless manifold is proportional to the manifold's Euler characteristic.\footnote{For manifolds with boundary, a boundary term needs to be added to the Lovelock action to produce the right Euler characteristic, see \eg \cite{Padmanabhan:2013xyr}. We review such term in the next section.} The variation of each of these topological terms can be written as a boundary term,\footnote{Indeed, locally, it is possible to write the terms `$\sqrt{|g|}\,\mathcal{L}_{D/2}$' themselves as total derivatives \cite{Padmanabhan:2013xyr}.} and does not contribute to the equations of motion of the Lovelock theory. This is the case \eg of Einstein gravity in two dimensions, and GB in four. However, the situation changes when the action is no longer a linear combination of ED like in the general $f($Lovelock$)$ theory. For example, terms of the form `$ \sqrt{|g|}\, R\cdot\mathcal{L}_2 $' are not topological in four dimensions. Another distinctive feature of Lovelock gravities which is not inherited in the more general $f($Lovelock$)$ framework is the fact that the former have second-order equations of motion. In fact, Lovelock gravities are the most general theories of gravity involving arbitrary combinations of the Riemann tensor which possess second-order equations of motion.\footnote{This statement is true for metric theories of gravity.} $f($Lovelock$)$ theories generically have fourth-order equations --- see \req{fLL1} and \req{fLL2}.

Certainly, the research area in which $f($Lovelock$)$ gravities have been considered more actively so far is cosmology --- see \eg \cite{DeLaurentis:2015fea,Jawad:2014kka,Atazadeh:2013cz,DeFelice:2010sh,Bamba:2009uf,Elizalde:2010jx,delaCruzDombriz:2011wn,Nojiri:2005jg,Nojiri:2006ri,Cognola:2006eg}, where, for example, they have been used to reproduce numerous features of the $\Lambda$-CDM model. In that context, the spacetime dimension is fixed to $D=4$ for obvious reasons, the $f($Lovelock$)$ action becomes a function of the Ricci scalar and the GB terms alone, and these theories are better known as `$f(R,\mathcal{G})$' gravities.

An interesting theoretical development was carried out in \cite{Sarkar:2013swa}. In that paper, the following formula for the gravitational entropy in $f($Lovelock$)$ theories was proposed,
\begin{equation}\label{SW1}
S_{\rm \ssc SW}=\frac{1}{4G}\int_m d^{(D-2)}x \sqrt{h_m}\, \sum_{p=1}^{\lfloor D/2 \rfloor}\, \left[p\, \frac{\partial f}{\partial \mathcal{L}_p} \cdot \prescript{(D-2)}{}{\mathcal{L}_{p-1}}\right] \, ,
\end{equation}
where $m$ is the corresponding horizon, and $\prescript{(D-2)}{}{\mathcal{L}_{p-1}}$ is the $(p-1)$-th ED associated to the pullback metric. This functional reduces to the well-known Jacobson-Myers functional (JM) for Lovelock gravities \cite{Jacobson:1993xs} and, as shown in \cite{Sarkar:2013swa}, it satisfies and increase theorem for small perturbations of Killing horizons as well as a generalized version of the second law for minimally-coupled fields. Apart from its interest in black hole thermodynamics, \req{SW1} has also been used in the holographic context. In fact, it is known \cite{deBoer:2011wk,Hung:2011xb} that the JM functional gives rise to the right universal terms when used to compute holographic entanglement entropy (HEE)\footnote{See section \ref{ee} for more details on entanglement entropy.} for these theories. This fact, along with the increase theorem already mentioned, was interpreted in \cite{Bueno1,Bueno2} as evidence for $S_{\rm \ssc SW}$ to be the right HEE functional for $f($Lovelock$)$ theories. The results found in those papers provide strong evidence that this is indeed the case.

The last two paragraphs summarize, to the best of our knowledge, the few aspects of general $f($Lovelock$)$ theories which have been so far studied in the literature. The goal of this paper is to develop several more.

\subsection{Main results}  \label{mr}
Our main results, section by section, can be summarized as follows:
\begin{itemize}
	\item In section \ref{var}, we generalize the Gibbons-Hawking-York (GHY) boundary term of general relativity \cite{York:1972sj, Gibbons:1976ue}, and its extensions to Lovelock \cite{Myers:1987yn,Teitelboim:1987zz} and $f(R)$ \cite{Madsen:1989rz} gravities to general $f($Lovelock$)$ theories. This new term --- see \req{bizcocho} below --- reduces to these in the appropriate subcases, and makes the $f($Lovelock$)$ action differentiable. The construction of this boundary term allows us to determine the number of physical degrees of freedom of the theory, which turns out to be $D(D-3)/2+r$, where $r$ is the rank of the Hessian matrix $H_{nm}=\partial_n\partial_m f$.
	\item In section \ref{eqe}, we make this counting of degrees of freedom explicit by showing that $f($Lovelock$)$ theories are equivalent to scalar-Lovelock gravities containing $r$ scalar fields.
	\item In section \ref{eq}, we linearize the $f($Lovelock$)$ equations on a maximally symmetric background (m.s.b.). Interestingly, we find that these theories do not propagate the usual ghost-like massive graviton characteristic of higher-derivative gravities. Furthermore, we show that certain non-trivial $f($Lovelock$)$ theories are also free of the --- also characteristic --- scalar mode, thus providing new examples of higher-derivative gravities which only propagate the usual physical graviton field on these backgrounds. For these theories, the equations of motion are second-order in any gauge, and the only effect of the higher-derivative terms appears in an overall factor whose effect is to change the normalization of the Newton constant. We provide examples of this class of theories in general dimensions.
	\item In section \ref{ee}, we consider holographic theories dual to some classes of $f($Lovelock$)$ theories and find constraints on the allowed values of their couplings. The first set of constraints is found by simply imposing the corresponding theory to admit an AdS$_D$ solution. After that we consider the holographic entanglement entropy of various entangling regions in the boundary theory, and find additional constraints by imposing the holographic surfaces to close off smoothly in the bulk.
	\item Last, but not least, in section \ref{sec:solutions} we construct new black hole solutions for certain $f($Lovelock$)$ theories. In particular, we start by embedding all solutions of pure Lovelock theory --- involving a single ED, $\mathcal{L}_n$, plus a cosmological constant --- in $f(\mathcal{L}_n)$, with special focus on static and spherically symmetric black holes. In particular, we construct the $f(\mathcal{L}_n)$ generalizations of the Schwarzschild(-AdS/dS) and Reissner-Nordstr\"om(-AdS/dS) black holes. We also construct new solutions for theories satisfying $f(\mathcal{L}^0_n)=f'(\mathcal{L}^0_n)=0$ for some constant $\mathcal{L}^0_n$. We go on to study under what conditions solutions of the general Lovelock theory can be embedded in $f($Lovelock$)$ theories depending on several ED. Finally, we construct a new static and spherically symmetric black hole solution of a particular $f(R,\mathcal{L}_2)$ theory in general dimensions.
	\item We comment on future directions in section \ref{disc}.
\end{itemize}
Let us get started.

\section{Variational problem and boundary term} 
\label{var}
In this section we study the variational problem in $f($Lovelock$)$ theories. Our main result is a new boundary term which generalizes the well-known GHY one for Einstein gravity as well as its generalizations to Lovelock and $f(R)$ theories. As we will see, the addition of this term to the $f($Lovelock$)$ action makes the corresponding variational problem well-posed.

A physical theory is often defined through an action functional, which is a map from a normed vector space (usually a space of functions) to the real numbers. On general grounds, the dynamical variables of the theory are described by some fields $\phi^a$. The action $S\left[\phi^a \right]$ consists in turn of a definite integral over a spacetime manifold $\mathcal{M}$, being the integrand a function of those fields and their derivatives, \ie
\begin{equation}\label{acgen}
S\left[\phi^a \right]=\int_{\mathcal{M}} d^Dx \sqrt{|g|} f(\phi^a, \nabla \phi^a,\dots) \, .
\end{equation}

Now, by a \emph{well-posed} variational problem we mean one for which the action functional
is differentiable. That is, under small variations of the fields $\phi^a \rightarrow \phi^a+\delta \phi^a$ we must be able to write the variation of the functional as
\begin{equation}
S\left[\phi^a+\delta\phi^a\right]-S\left[\phi^a \right]=\delta S\left[\phi^a,\delta\phi^a\right]+\mathcal{O}\left({(\delta\phi^a)}^2\right) \, ,
\end{equation}
where $\delta S\left[\phi^a,\delta\phi^a\right]$ is linear on $\delta\phi^a$. If we perform this variation explicitly in \req{acgen}, we find two terms, namely
\begin{equation}
\delta S = \int_{\mathcal{M}} d^Dx \sqrt{|g|}\, \mathcal{E}_a\, \delta\phi^a + \int_{\partial\mathcal{M}} d^{D-1}x \sqrt{\vert h \vert}\, \theta \left(\phi^a,\nabla\phi^a,\delta\phi^a,\nabla\delta\phi^a,\dots \right) \, .
\end{equation} 
Here, $\mathcal{E}_a$ is a function of the fields and their derivatives, $h$ is the determinant of the induced metric on the boundary $\partial \mathcal{M}$ and $\theta$ is some function of the fields and their derivatives. While the first term is linear on $\delta\phi^a$, the second is not necessarily of that form. Field perturbations need to respect the field (Dirichlet) boundary conditions, \ie they are required to satisfy $\delta\phi^a\vert_{\partial\mathcal{M}}=0$. However in general $\nabla\delta\phi^a\vert_{\partial\mathcal{M}}\neq 0$ and in consequence this boundary term may not be trivially zero, making the action functional non-differentiable. 

When this is the case, one can sometimes modify the original action by introducing an appropriate boundary term such that its variation cancels this contribution. When it can be constructed, this boundary term makes the functional differentiable and the variational problem becomes well-posed. After the addition of the boundary term, the complete action reads
\begin{equation}
S\left[\phi^a \right]=\int_{\mathcal{M}} d^Dx \sqrt{|g|}\, f(\phi^a, \nabla \phi^a,\dots)+ \int_{\partial\mathcal{M}} d^{D-1}x \sqrt{\vert h \vert}\, \psi \left(\phi^a,\nabla\phi^a,\dots \right) \, .
\end{equation}
Now, its variation is simply given by
\begin{equation}
\delta S = \int_{\mathcal{M}} d^Dx \sqrt{|g|}\, \mathcal{E}_a\, \delta\phi^a  \, ,
\end{equation} 
because $\psi$ has been chosen in a way such that $ \left( \theta +\delta\psi \right)=0$. 

The principle of least action asserts that a field configuration $\phi^a_0$ is a solution of the theory if it constitutes a stationary point of the action functional, \ie if $\delta S\left[\phi^a_0\right]=0$. Hence, solutions of the theory satisfy the equations of motion $\mathcal{E}_a=0$.

Before we go on, let us mention that, in general, the boundary term is not the only addition to the original action that needs to be made. In particular, extra counter-terms usually need to be included in order for the action to be finite when evaluated on configurations satisfying the equations of motion. We will not be concerned with that issue here.\footnote{Let us parenthetically mention that such counter-terms where constructed for AdS$_D$ spacetimes in \cite{Emparan:1999pm} and \cite{Yale:2011dq,Kofinas:2007ns} for Einstein and Lovelock theories respectively.}

\subsection{Equations of motion}
Let us now see how the ideas sketched in the previous subsection apply to the $f($Lovelock$)$ theory, whose action is given by \req{flov}. If we vary this action with respect to the metric, we find
\begin{equation}
\delta S_{ f(\rm{Love.})}=\frac{1}{16\pi G}\int_{\mathcal{M}} d^Dx\sqrt{|g|}\,\E_{\mu\nu}\delta g^{\mu\nu}+\frac{\varepsilon}{16\pi G}\int_{\partial \mathcal{M}} d^{D-1}x\sqrt{|h|}\sum_{n=1}^{\lfloor D/2 \rfloor} \delta v^{\mu}_{n}n_{\mu}\partial_n f\, ,
\label{varf}
\end{equation}
where we have used the Stokes theorem in the second term. Here,  $n_{\mu}$ is a vector orthonormal to the boundary $\partial \mathcal{M}$ with $n_{\mu}n^{\mu}=\varepsilon$
and $h_{\mu\nu}=g_{\mu\nu}-\varepsilon\, n_{\mu}n_{\nu}$ is the pullback metric. 
The quantities $\E_{\mu\nu}$ and $\delta v^{\mu}_{j}$ are given by
\begin{equation}
\E_{\mu\nu}=\sum_{n=1}^{\lfloor D/2 \rfloor}\left[\mathcal{E}_{\mu\nu}^{(n)}+\frac{1}{2}g_{\mu\nu}\mathcal{L}_{n}-2P^{(n)}_{\alpha\nu\lambda\mu}\nabla^{\alpha}\nabla^{\lambda}\right]\partial_{n}{f}-\frac{1}{2}g_{\mu\nu}f\, ,\label{fLL1}
\end{equation}
and
\begin{equation}
\delta v^{\mu}_{j}=2g^{\beta\sigma} P^{(j)\mu\nu}_{\alpha\beta} \nabla^{\alpha}\delta g_{\nu\sigma}\, ,
\end{equation}
respectively. In these expressions we have defined the following tensors\footnote{Both tensors are divergence-free in all indices, \ie $\nabla^{\mu} \mathcal{E}^{(j)}_{\mu\nu}=0$, $\nabla^{\alpha} P^{(j)\mu\nu}_{\alpha\beta}=0$.}
\begin{equation}
\mathcal{E}_{\mu\nu}^{(j)}=\frac{-1}{2^{j+1}}g_{\alpha \mu}\delta^{\alpha \mu_1\dots \mu_{2j}}_{\nu \nu_1\dots \nu_{2j}} R^{\nu_1\nu_2}_{\mu_1\mu_2}\cdots R^{\nu_{2j-1}\nu_{2j}}_{\mu_{2j-1}\mu_{2j}}\, , \quad P^{(j)\mu\nu}_{\alpha\beta}=\frac{-j}{2^j}\delta^{\mu\nu \sigma_1\dots\sigma_{2j-2}}_{\alpha \beta \lambda_1\dots \lambda_{2j-2}}R_{\sigma_1\sigma_2}^{\lambda_1\lambda_2}\cdots R_{\sigma_{2j-3}\sigma_{2j-2}}^{\lambda_{2j-3}\lambda_{2j-2}} \, .
\label{love}
\end{equation}
Also, in \req{varf} and \req{fLL1} we have used the notation $\partial_{n}f= \partial f/\partial \mathcal{L}_{n}$, which will appear throughout the text. Now, if we forget the boundary contribution for a moment, we see that the equations of motion of the theory read
\begin{equation}
\E_{\mu\nu}=0\, ,\label{fLL2}
\end{equation}
whose trace is\footnote{
In order to get this result we used the relations: 
	$
	\E^{(n)\alpha}_{\alpha}=(n-D/2)\mathcal{L}_{n}$ and $ P^{(n)\lambda\mu}_{\alpha\mu}=n(D-2n+1)\E^{(n-1)\lambda}_{\alpha}\, .
	$
}
\begin{equation}
\sum_{n=1}^{\lfloor D/2 \rfloor}\left[n\mathcal{L}_{n}-2n(D-2n+1)\E_{\mu\nu}^{(n-1)}\nabla^{\mu}\nabla^{\nu}\right]\partial_n f-\frac{D}{2}f=0.
\end{equation}
As expected, these reduce to the Lovelock and $f(R)$ equations of motion, 
\begin{equation}\label{efl}
\sum_{n=0}^{\lfloor D/2 \rfloor}\lambda_{n}\,\Lambda_0^{1-n}\mathcal{E}_{\mu\nu}^{(n)}=0\, ,�\quad \mathcal{E}_{\mu\nu}^{f(R)}= f'(R)R_{\mu\nu}-\frac{1}{2}f(R)g_{\mu\nu}+\left(g_{\mu\nu}\Box-\nabla_{\mu}\nabla_{\nu}\right)f'(R)=0\, ,
\end{equation}
when we choose $f=f_{\rm Love.}$ and $f=f_{f(R)}$ as in \req{cases} respectively. In particular, observe that $\nabla^{\lambda}(\partial_j f)=0$ $\forall\, j$ when $f$ is a linear combination of ED --- corresponding to the usual Lovelock theory --- so the term contributing with fourth-order derivatives in \reef{fLL1} disappears in that case. In appendix \ref{frl2}, we provide the explicit equations of motion corresponding to $D$-dimensional $f($Lovelock$)$ theories which are only functions of the Ricci scalar and the GB terms, \ie $f=f(R,\mathcal{L}_2)$. These are, in particular, the most general $f($Lovelock$)$ gravities in four dimensions, as the densities $\mathcal{L}_p$ identically vanish for all $p\geq 3$ in that case.

\subsection{Generalized boundary term}
Let us now see what happens with the boundary contribution to $\delta S$. As we explained at the beginning of this section, the variational problem for \req{flov} cannot be well-posed because such an action is not differentiable, as is clear from the presence of the boundary term in \req{varf}.

In the familiar case of Einstein gravity, the problem is solved through the introduction of the usual GHY term \cite{York:1972sj, Gibbons:1976ue}
\begin{equation}
S_{\rm EH}\rightarrow S_{\rm EH}+S_{\rm GHY}\,, \quad \text{where}\quad S_{\rm GHY}=\frac{\varepsilon}{8\pi G}\int_{\partial \mathcal{M}}d^{D-1}x\sqrt{|h|}K,
\end{equation}
and $K$ is the trace of the second fundamental form associated to the boundary normal $n_{\mu}$, \ie $K= g^{\mu\nu} K_{\mu\nu}$, where $K_{\mu\nu}=h_{\mu}^{\rho}\nabla_{\rho}n_{\nu}$. It is a standard exercise to show that the variation of this corrected action does not contain additional boundary terms as long as the usual Dirichlet boundary condition\footnote{Clearly, boundary terms in general, and the GHY one in particular, are not unique. They are only unique up to contributions whose variations vanish when we impose Dirichlet boundary conditions.}
\begin{equation}\label{bdyc2}
\delta g_{\mu\nu}\Big|_{\partial \mathcal{M}}=0\, ,
\end{equation}
is satisfied. Indeed, the variation of $K$ produces a term which exactly cancels the original boundary contribution coming from the variation of the EH action, plus additional terms which vanish for configurations respecting \req{bdyc2}. Hence, the corrected EH action is differentiable. Since the only condition we need to impose in order to find a solution to the theory is \req{bdyc2}, \ie we only need to fix the metric at the boundary, we can obtain the number of classical degrees of freedom of Einstein gravity as the number of independent components of the boundary metric. This yields the well-known result: $n_{\rm dof}=D(D-3)/2$.

The problem becomes more involved in Lovelock and $f(R)$ gravities, for different reasons in each case. 
Lovelock theories possess second-order equations of motion, and the metric does not propagate additional degrees of freedom with respect to Einstein gravity. Therefore, the only boundary condition that one needs to impose is again given by \req{bdyc2}. However, the boundary term that needs to be added to the usual Lovelock action --- see \req{flov} and \req{cases} --- in order to make it differentiable is considerably more involved than the GHY term. The full Lovelock action is given by\footnote{The `MTZ' label here stands for Myers \cite{Myers:1987yn}, Teitelboim and Zanelli \cite{Teitelboim:1987zz} who independently first showed how to construct this boundary term. The equivalence of both approaches was proven in \cite{Miskovic:2007mg}.}
\begin{equation}\label{LLactionfull}
S_{\rm Love.}\rightarrow S_{\rm Love.}+S_{\rm MTZ}\,, \quad \text{where}\quad S_{\rm MTZ}=\frac{\varepsilon}{16\pi G}\sum_{n=0}^{\lfloor D/2 \rfloor}\lambda_n \Lambda_0^{1-n}\, \int_{\partial \mathcal{M}}d^{D-1}x\sqrt{|h|}Q_{n}\, ,
\end{equation}
and where
\begin{equation}
Q_{n}=2n\int_0^1dt\, \delta^{\mu_1\dots\mu_{2n-1}}_{\nu_1\dots\nu_{2n-1}}K^{\nu_1}_{\mu_1}\left[\frac{1}{2}R^{\nu_2\nu_3}_{\mu_2\mu_3}-t^2K^{\nu_2}_{\mu_2}K^{\nu_3}_{\mu_3}\right]\cdots \left[\frac{1}{2}R^{\nu_{2n-2}\nu_{2n-1}}_{\mu_{2n-2}\mu_{2n-1}}-t^2K^{\nu_{2n-2}}_{\mu_{2n-2}}K^{\nu_{2n-1}}_{\mu_{2n-1}}\right]\, .
\label{Qterm}
\end{equation}
Indeed, it is possible to prove that 
\begin{equation}
\delta Q_{n}\Big|_{\delta g_{\mu\nu}|_{\partial \mathcal{M}}=0}=n_{\mu}\delta v^{\mu}_{n} \, ,
\end{equation}
\ie the variation of this term exactly cancels the boundary contribution which appears from the variation of $S_{\rm Love.}$ as long as the boundary condition \req{bdyc2} is satisfied. Therefore, the addition of $S_{\rm MTZ}$ makes the Lovelock variational problem well-posed. Of course, $S_{\rm MTZ}$ reduces to $S_{\rm GHY}$ in the particular case of Einstein gravity. 

As opposed to Lovelock theories, general $f(R)$ gravities have fourth-order equations of motion. This means that the theory contains more degrees of freedom than Lovelock gravity and that besides \req{bdyc2}, additional boundary conditions must be imposed. In fact, as we review in section \ref{eqe}, $f(R)$ gravities with $f^{\prime\prime}(R)\neq 0$ are equivalent to Brans-Dicke theories, in which a scalar field $\phi$ related to the $f(R)$ metric through $\phi=f'(R)$ is coupled to the gravitational field. Hence, it is natural to expect that a condition of the form 
\begin{equation}\label{condimento}
\delta \phi |_{\partial \mathcal{M}}=\delta(f'(R))|_{\partial \mathcal{M}}=(f''(R)\delta R)|_{\partial \mathcal{M}}=0 \rightarrow \delta R|_{\partial \mathcal{M}}=0 \, ,
\end{equation}
needs to be added in that case. On the other hand, we expect again the boundary term to reduce to the GHY one for $f(R)=R-2\Lambda_0$. These observations turn out to be right, as the $f(R)$ variational problem can be made well-posed by considering the following action\footnote{In this case, the first to have considered this boundary term seem to have been Madsen and Barrow in \cite{Madsen:1989rz}. See also \cite{Guarnizo:2010xr,Dyer:2008hb}.}
\begin{equation}
S_{ f(R)}\rightarrow S_{f(R)}+S_{\rm MB}\,, \quad \text{where}\quad S_{\rm MB}=\frac{\varepsilon}{8\pi G}\int_{\partial \mathcal{M}}d^{D-1}x\sqrt{|h|}f'(R)K \, .
\label{fRfull}
\end{equation}
This trivially reduces to $S_{\rm GHY}$ for Einstein gravity. Besides, its variation precisely compensates the extra boundary term produced from the variation of $S_{ f(R)}$. In particular, imposing \req{bdyc2} one finds
\begin{equation}
\delta S_{f(R)}+\delta S_{\rm MB}=\frac{1}{16\pi G}\int_{\mathcal{M}}d^{D}x \sqrt{|g|}\, \mathcal{E}_{\mu\nu}^{f(R)}\, \delta g^{\mu\nu}+ \frac{\varepsilon}{8\pi G}\int_{\partial \mathcal{M}}d^{D-1}x\sqrt{|h|} K\, \delta (f'(R)) \, ,
\label{fRfullvara}
\end{equation}
where $\mathcal{E}_{\mu\nu}^{f(R)}$ is given in \req{efl}. Hence, we observe that imposing the additional boundary condition \req{condimento} on $f'(R)$ --- or equivalently, on $R$ --- makes the corrected action differentiable. While one might feel uncomfortable at first by imposing boundary conditions on functions that depend on derivatives of the metric like \req{condimento}, let us stress that the introduction of $S_{\rm MB}$ is necessary to reproduce the correct ADM energy in the Hamiltonian formalism as well as the right black hole entropy --- \ie one which matches the result obtained with Wald's formula --- using the Euclidean semiclassical approach \cite{Dyer:2008hb}.
We observe that $f(R)$ theories with $f^{\prime\prime}(R)\neq 0$ have $D(D-3)/2+1$ degrees of freedom.

Let us finally turn to the general $f($Lovelock$)$ case. We propose the following boundary term 
\begin{equation}\label{bizcocho}
S_{ f(\rm{Love.})}\rightarrow S_{ f(\rm{Love.})}+\tilde{S}\, , \quad \text{where}\quad \tilde{S}=\frac{\varepsilon}{16\pi G}\int_{\partial \mathcal{M}}d^{D-1}x\sqrt{|h|}\sum_{n=1}^{\lfloor D/2 \rfloor}\partial_n f(\mathcal{L}) Q_{n}\, ,
\end{equation}
and where $S_{ f(\rm{Love.})}$ is given in \req{flov}. It is straightforward to check that this reduces to $S_{\rm MB}$, $S_{\rm MTZ}$ and $S_{\rm GHY}$ in the particular cases of $f(R)$, Lovelock and Einstein gravity respectively. After imposing the Dirichlet condition \req{bdyc2} on the boundary metric, the variation of this corrected action becomes
\begin{equation}\label{varflov}
\delta S_{ f(\rm{Love.})}+\delta\tilde{S}=\frac{1}{16\pi G}\int_{\mathcal M} d^Dx\sqrt{|g|}\, \E_{\mu\nu}\, \delta g^{\mu\nu}+\frac{\varepsilon}{16\pi G}\int_{\partial \mathcal{M}} d^{D-1}x\sqrt{|h|}\sum_{n,m=1}^{\lfloor D/2 \rfloor} \partial_m\partial_n f \, \delta \mathcal{L}_{m} Q_{n}\, ,
\end{equation}
where we have used the relation
\begin{equation}
\delta (\partial_n f )=\sum_{m=1}^{\lfloor D/2 \rfloor} \, \partial_m\partial_n f \, \delta \mathcal{L}_{m}\, .
\end{equation}
Equation \req{varflov} suggests that, in addition to the metric, we need to fix the Euler densities at the boundary, \ie
\begin{equation}\label{cond1}
 \delta \mathcal{L}_{n}\Big|_{\partial \mathcal{M}}=0,\quad  n=1,\dots,\lfloor D/2 \rfloor.
\end{equation}
However, notice that it is enough to fix the derivatives of $f$,
\begin{equation}
\delta \left(\partial_n f\right)\Big|_{\partial \mathcal{M}}=0,\quad  n=1,\dots,\lfloor D/2 \rfloor,
\label{Bcond}
\end{equation}
which is a weaker condition in general. If the Hessian matrix, $H_{nm}=\partial_n\partial_m f$,  is not singular, \ie if $\displaystyle \det H_{nm}\neq 0$, then the conditions (\ref{Bcond}) and \req{cond1} are equivalent. But if this determinant is zero, then not all the conditions in (\ref{Bcond}) are independent. In fact, if $r$ is the rank of the Hessian matrix, 
\begin{equation}
r= \displaystyle\operatorname{rank}H_{nm}\, ,
\end{equation}
then there are $r$ independent conditions. Thus, only $r$ quantities must be fixed at the boundary and the number of physical degrees of freedom in $f$(Lovelock) theory is given by:
\begin{equation}
n_{{\rm dof}}=\frac{D(D-3)}{2}+r.
\end{equation}
With respect to GR or Lovelock gravity there are $r$ additional degrees of freedom. Depending on the function, $r$ can take values from $0$ to $\lfloor D/2 \rfloor$. In the next section we will see that these additional degrees of freedom can be interpreted as scalar fields in an equivalent scalar-Lovelock theory.

\section{Equivalence with scalar-tensor theories} 
\labell{eqe}
It is a well known fact that $f(R)$ gravity is equivalent to a scalar-tensor theory of the Brans-Dicke class --- see \eg \cite{Wands:1993uu,Sotiriou:2008rp}. This can be easily seen by considering an action of the form
\begin{equation}\label{bd}
S=\frac{1}{16\pi G}\int_{\mathcal{M}} d^Dx\sqrt{|g|}\, \left[f(\chi)+f^{\prime}(\chi)(R-\chi)\,  \right]\, .
\end{equation}
The equation of motion for the auxiliary field $\chi$, $f^{\prime\prime}(R)(R-\chi)=0$, implies $\chi=R$ provided $f^{\prime\prime}(R)\neq 0$. Substituting this back in \req{bd}, we recover the $f(R)$ action. Now, assuming the field redefinition $\phi=f^{\prime}(\chi)$ can be inverted,\footnote{A sufficient condition for this is $f^{\prime\prime}(R)\neq 0$.} we can rewrite \req{bd} as
\begin{equation}\label{bed}
S_{\rm BD}=\frac{1}{16\pi G}\int_{\mathcal{M}} d^Dx\sqrt{|g|}\, \left[\phi\,R-V(\phi)  \right]\, ,
\end{equation}
where $V(\phi)=\chi(\phi)\phi-f(\chi(\phi))$. This is the action of a Brans-Dicke theory with parameter $\omega_0=0$.

The situation is slightly more sophisticated in the case of $f($Lovelock$)$ theories. In analogy with \req{bd}, let us consider the following action containing $\lfloor D/2 \rfloor$ auxiliary scalar fields $\chi_1,\dots,\chi_{\lfloor D/2 \rfloor}$,
\begin{equation}\label{bdlove}
S=\frac{1}{16\pi G}\int_{\mathcal{M}} d^Dx\sqrt{|g|}\, \left[f+\sum_{n=1}^{\lfloor D/2 \rfloor} \partial_n f\cdot(\mathcal{L}_n-\chi_n)\,  \right]\, ,
\end{equation}
where $f=f(\chi_1,\dots,\chi_{\lfloor D/2 \rfloor})$. The equations of motion for the auxiliary fields are constraints which relate them to the dimensionally-extended Euler densities,
\begin{eqnarray}
\label{phiLL}
\sum_{n=1}^{\lfloor D/2 \rfloor}\partial_n\partial_m f\cdot\left(\mathcal{L}_{n}-\chi_n\right)=0, \quad m=1,\dots,\lfloor D/2 \rfloor\, .
\label{phiconst}
\end{eqnarray}
Hence, we see that if we set 
\begin{eqnarray}\label{lnphi}
\mathcal{L}_{n}=\chi_n\, ,\,\,\, n=1,\dots,\lfloor D/2 \rfloor\, ,
\end{eqnarray}
 \req{phiLL} is satisfied and \req{bdlove} reduces to the $f($Lovelock$)$ action \req{flov}. In general, however, this will not be the only solution to \req{phiLL}. There are two possibilities that we explain in the following subsections.

\subsection{Non-degenerate case}
If the Hessian matrix $H_{nm}$ is non-singular, \ie if $\det (H_{nm})\neq 0$, \req{lnphi} is indeed the only solution to the constraint equations \req{phiLL}, and the action \req{bdlove} is equivalent to the original $f($Lovelock$)$ one \req{flov}.

In this situation, we can perform the invertible field redefinition
\begin{eqnarray}\label{lnphi2}
\phi_n=\partial_n f(\chi_1,\dots,\chi_{\lfloor D/2 \rfloor})\, , \,\,\, n=1,\dots,\lfloor D/2\rfloor\, ,
\end{eqnarray}
which allows us to rewrite \req{bdlove} as
\begin{equation}\label{bed3}
S=\frac{1}{16\pi G}\int_{\mathcal{M}} d^Dx\sqrt{|g|}\, \left[\sum_{n=1}^{\lfloor D/2\rfloor}\phi_n \mathcal{L}_n\,-V  \right]\, ,
\end{equation}
where\footnote{We use the notation `$\chi$' and `$\phi$' to generically refer to the $\lfloor D/2\rfloor$ scalars $\chi_n$ and the same number of $\phi_n$. For example, $f(\chi(\phi))$ stands for $f(\chi_1(\phi_1,\dots,\phi_{\lfloor D/2\rfloor}),\dots,\chi_{\lfloor D/2 \rfloor}(\phi_1,\dots,\phi_{\lfloor D/2\rfloor}))$.}
\begin{equation}\label{bed2}
V=\sum_{n=1}^{\lfloor D/2\rfloor} \chi_n(\phi) \phi_n-f(\chi(\phi))\, 
\end{equation}
is the Legendre transform of $f$. This form of the action, which clearly resembles --- and generalizes --- the $f(R)$ scalar-tensor action in \req{bed}, was first noted to be related to the $f($Lovelock$)$ action \req{flov} in \cite{Sarkar:2013swa}.

\subsection{Degenerate case}
If the Hessian matrix is singular, \ie if $\det (H_{nm})=0$, the system of equations (\ref{phiconst}) is indeterminate. In particular, the space of solutions has dimension $\lfloor D/2 \rfloor-r$, where $r= \operatorname{rank}(H_{mn})$. Hence, unless $r=\lfloor D/2 \rfloor$, which corresponds to the case studied in the previous subsection\footnote{Indeed, if $r=\lfloor D/2 \rfloor$ then $\det (H_{nm})\neq 0$ and the dimensionality of the space of solutions is $0$, \ie there is a unique solution given by \req{lnphi}.}, there are infinite solutions to \req{phiconst}. This is nothing but a manifestation of the fact that we have included $\lfloor D/2 \rfloor-r$ too many scalars to account for the actual number of physical degrees of freedom of the corresponding $f($Lovelock$)$ theory. Let us see how we can reduce this number. It is clear that we cannot perform a Legendre transform this time, because the Hessian matrix is singular, which implies that the change of variables in \req{lnphi2} is not invertible. We can make, however, a \emph{semi}--Legendre transform. This goes as follows: let us define the fields $\phi_n$ as before:
\begin{equation}
\phi_{n}=\partial_n f(\chi_1,\dots,\chi_{\lfloor D/2 \rfloor}), \quad n=1,\dots,\lfloor D/2 \rfloor.
\end{equation}
Then, there is a subset $I\subset \{1,\dots,\lfloor D/2 \rfloor\}$ of $r$ indices such that $\phi_I= \{\phi_i\}_{i\in I}$ are independent variables, in the sense that
\begin{equation}
\operatorname{det}\left[(\partial \phi_{i_1}/\partial \chi_{i_2})_{i_1,i_2\in I}\right]=\operatorname{det}\left[(H_{{i_1}i_2})_{i_1,i_2\in I}\right]\neq 0.
\end{equation}
Let $J$ be the complementary set of indices, $J=\{1,\dots,\lfloor D/2 \rfloor\}-I$.  Now, since there must be only $r$ independent fields, the rest of the fields, $\phi_j, j\in J$ must depend on the formers $\phi_I$. Hence, there exist some functions $g_j$ such that 
\begin{equation}
\phi_j=g_j(\phi_{I}), \quad j\in J.
\end{equation}
Then, we can consider $\phi_I\cup\chi_J$ as our set of \emph{independent} variables.\footnote{The change of variables $(\chi_n)\rightarrow (\phi_I; \chi_J)$ is now invertible.} 
We define the \emph{semi}--Legendre transform of $f$ as:
\begin{equation}
\tilde V(\phi_{I})=\sum_{i\in I}\chi_i(\phi_I, \chi_J)\cdot \phi_i+\sum_{j\in J}\chi_j\cdot g_j(\phi_I)-f\left(\chi_I(\phi_I, \chi_J);\chi_J\right).
\end{equation}
This seems to be a function of both the $\phi_I$ and the $\chi_J$. However, it is easy to check that the derivative of $\tilde{V}$ with respect to any $\chi_J$ vanishes, $\partial_J \tilde V=0$, which implies that $\tilde{V}$ is actually a function of the $r$ fields $\phi_I$ alone. This allows us to rewrite the original action \req{bdlove} as \begin{equation}\label{dege}
S=\frac{1}{16\pi G}\int_{M}d^Dx\sqrt{|g|}\left[\sum_{i\in I}\phi_i\mathcal{L}_{i}+\sum_{j\in J}g_j(\phi_I)\mathcal{L}_{j}-\tilde V(\phi_I) \right].
\end{equation}
This theory is equivalent to $f($Lovelock$)$, since we have eliminated the spurious degrees of freedom that appeared in the original action (\ref{bdlove}). The equations of motion for the scalar fields have now a unique solution given precisely by $\chi_n(\phi_I)=\mathcal{L}_n$. 

We see that, on general grounds, $f($Lovelock$)$ gravity is equivalent to a scalar-Lovelock theory with $r$ scalars, where $r$ is the rank of the Hessian matrix of $f$, and whose action is given by \req{dege}. In the case of Lovelock gravity such an analogy does not exist: the Hessian is zero and so is the number of scalars. In appendix \ref{examples}, we explicitly construct the equivalent scalar-Lovelock theories for a pair of classes of $f($Lovelock$)$ theories including both degenerate and non-degenerate subcases.

Let us finally mention that Lovelock theories have been recently proposed to be effectively described by Einstein gravity coupled to certain $p$-form gauge fields \cite{Brustein:2012uu}.\footnote{See also \cite{Cuzinatto:2006hb}.} We will not explore here how such relation might extend to the more general $f($Lovelock$)$ scenario.


\section{Linearized equations of motion} 
\labell{eq}
In this section we study the linearized equations of motion of $f($Lovelock$)$ gravity on a general m.s.b., with particular emphasis on AdS$_D$. On general grounds, the linearized equations of motion of higher-derivative gravities on a m.s.b. are fourth-order in derivatives. From these equations it is possible to identify, in addition to the usual spin-2 massless graviton, a scalar field corresponding to the trace of the perturbation, as well as an additional massive spin-2 field, which generally presents an undesirable ghost-like behavior --- see \eg \cite{Myers:2010tj} for a discussion. Remarkably, we find that for general $f($Lovelock$)$ gravities, this massive graviton is absent, and the linearized equations of motion are second-order. Further, we find that for certain non-trivial classes of theories, the extra spin-0 degree of freedom is also absent, hence providing examples of theories for which, just like for Einstein or Lovelock, the only dynamical perturbation on a m.s.b. is the usual massless graviton.

An interesting motivation for constructing higher-derivative theories without the extra spin-0 and spin-2 modes in AdS$_D$ backgrounds was made clear in \cite{Myers:2010ru}, where the authors observed that a particular higher-derivative theory containing a non-trivial cubic term \cite{Oliva:2010eb,Myers:2010ru} --- and which is well-known now as quasi-topological gravity\footnote{Higher-derivative extensions of quasi-topological gravity were constructed in \cite{Dehghani:2011vu,Oliva:2011xu}.} --- was also free of these extra fields.\footnote{Indeed, the linearized equations of quasi-topological gravity are identical to those of Einstein gravity up to an overall factor \cite{Myers:2010ru}.} The reason is that holographic calculations involving graviton propagators become easily doable for theories satisfying this property, while providing non-trivial information about the dual CFTs. In the case of quasi-topological gravity, these holographic studies were performed in \cite{Myers:2010jv}.

\subsubsection*{Maximally symmetric solutions}
Let us consider the following action,
\begin{equation}
S=\frac{1}{16\pi G}\int_{\mathcal{M}}d^Dx\sqrt{|g|}\left[-2\Lambda_0+R+\lambda  f(\mathcal{L}_0,\mathcal{L}_1,\dots,\mathcal{L}_{\lfloor D/2 \rfloor})\right]\, ,
\label{paramf}
\end{equation}
\ie we make explicit the EH and cosmological constant terms for clarity reasons. The equation of motion is simply given by
\begin{equation}
R_{\mu\nu}-\frac{1}{2}g_{\mu\nu}R+\Lambda_0\, g_{\mu\nu}+\lambda\,\E_{\mu\nu}=0\, ,\label{fLL3}
\end{equation}
where $\E_{\mu\nu}$ is given in \req{fLL1}.
As anticipated, we assume our background to be maximally symmetric, with metric $\bar g_{\mu\nu}$. The Riemann tensor of such spacetime is given by
\begin{equation}
\bar R^{\mu\nu}_{\alpha\beta}=\Lambda \delta^{\mu\nu}_{\alpha\beta},
\label{LambdaRiemann}
\end{equation} 
for some real constant $\Lambda$ which, for AdS$_D$, is related to the AdS radius $L$ by $\Lambda=-1/ L^2$. When $\lambda=0$, \ie for Einstein gravity, the maximally symmetric spacetime satisfying \req{LambdaRiemann} is a solution of the theory provided $\Lambda$ is related to the cosmological constant $\Lambda_0$ through
\begin{equation}
\Lambda=\frac{2\Lambda_0}{(D-1)(D-2)}\, .
\end{equation}
For non-vanishing $\lambda$, we find the following constraint equation


\begin{equation}
\begin{aligned}
\frac{2\Lambda_0-\lambda\,  f\left(\bar{\mathcal{L}}\right)}{(D-1)(D-2)}=\Lambda-\sum_{n=1}^{\lfloor D/2\rfloor}\frac{2\lambda \,n(D-3)!}{(D-2n)!}\Lambda^n\,\partial_n f\left(\bar{\mathcal{L}}\right)\, ,
\label{lambdaconst}
\end{aligned}
\end{equation}
where the bars mean that the corresponding quantities are evaluated on the background metric, and where we have used the following expressions
\begin{eqnarray}
\bar{\mathcal{L}}_{n}=\frac{D!}{(D-2n)!}\Lambda^n\, ,\quad
\bar \E_{\mu\nu}^{(n)}=-\frac{1}{2}\frac{(D-1)!}{(D-2n-1)!}\Lambda^n g_{\mu\nu}\, .
\end{eqnarray}
Given $f$ and $\Lambda_0$, \req{lambdaconst} is an algebraic equation for $\Lambda$, and its solutions determine the possible vacua of the theory. In general, some of these vacua will contain ghost-like gravitons and will be unstable \cite{Boulware:1985wk}. Note that this can occur even if the theory propagates only a single graviton mode,\footnote{We thank Rob Myers for clarifying this point to us.} like in the case of Lovelock theories. This problem can be avoided by choosing the vacuum that reduces to the Einstein gravity one when the higher-order couplings vanish --- \ie when $\lambda\rightarrow 0$ in the case considered here.

As we will see in section \ref{ee}, the embedding equation \req{lambdaconst} can be used to constrain the space of allowed values for the $f($Lovelock$)$ couplings.

\subsubsection*{Linearized equations}
Let us now consider a small perturbation of our background metric, $g_{\mu\nu}=\bar g_{\mu\nu}+h_{\mu\nu}$, with $h_{\mu\nu}<<1$ for all $\mu,\nu=0,\dots,D$. At linear order in $h_{\mu\nu}$, the ED and the tensors $\E_{\mu\nu}^{(n)}$ read
\begin{eqnarray}
\label{Llinear}
\mathcal{L}_{n}=\bar{\mathcal{L}}_{n}+\frac{n(D-2)!}{(D-2n)!}\Lambda^{n-1}R^L\, , \quad 
\E_{\mu\nu}^{(n)}=\bar \E_{\mu\nu}^{(n)}+\frac{n(D-3)!}{(D-2n-1)!}\Lambda^{n-1}G^L_{\mu\nu}\, ,
\end{eqnarray}
where $R^L$ and $G^{L\mu}_{\nu}$ are the linearized Ricci scalar and Einstein tensor respectively.\footnote{The linearized Einstein tensor, Ricci tensor and Ricci scalar are given respectively by 
	\begin{eqnarray}
	G_{\mu}^{L\alpha}&=& \bar{g}^{\alpha\nu}R_{\mu\nu}^{L}-\frac{1}{2}\delta_{\mu}^{\alpha}R^L-\Lambda(D-1)h_{\mu}^{\alpha}\, . \\
	R^{L}_{\mu\nu}&=&\frac{1}{2}\bar\nabla_{\mu}\bar\nabla_{\sigma} h^{\sigma}_{\nu}+\frac{1}{2}\bar\nabla_{\nu}\bar\nabla_{\sigma}h^{\sigma}_{\mu}-\frac{1}{2}\bar\nabla_{\nu}\bar\nabla_{\mu}h-\frac{1}{2}\bar\Box h_{\mu\nu}+\Lambda D h_{\mu\nu}-\Lambda h \bar g_{\mu\nu}\, ,\\
	R^L&=&\bar\nabla_{\mu}\bar\nabla_{\nu}h^{\mu\nu}-\bar\Box h-\Lambda (D-1)h\, .
	\end{eqnarray}}
The following result is also necessary: 
\begin{equation}
\bar P^{(n)\mu\nu}_{\alpha\beta}=-\frac{n(D-2)!}{2(D-2n)!}\Lambda^{n-1}\delta^{\mu\nu}_{\alpha\beta}.
\label{Plinear}
\end{equation}
Using this information we can find the linearized version of \req{fLL3}. It reads
\begin{equation}\label{lini}
\alpha \,G^L_{\mu\nu}+\Lambda \,\beta\,\bar{g}_{\mu\nu}R^L+\frac{\beta}{(D-1)}(\bar{g}_{\mu\nu}\bar{\Box}-\bar\nabla_{\nu}\bar\nabla_{\mu})R^L=0\, ,
\end{equation}
where $\alpha$ and $\beta$ are the following constants
\begin{eqnarray}
\label{alpha}
\alpha&=&1+\lambda\sum_{n=1}^{\lfloor D/2 \rfloor}\partial_n f\left(\bar{\mathcal{L}}\right)\frac{n(D-3)!}{(D-2n-1)!}\,\Lambda^{n-1},\\
\beta&=&\lambda \sum_{n,m=1}^{\lfloor D/2 \rfloor}\partial_n\partial_m f\left(\bar{\mathcal{L}}\right) \frac{n\,m(D-2)!(D-1)!}{(D-2n)!(D-2m)!}\,\Lambda^{n+m-2}.
\label{beta}
\end{eqnarray}
Note that these constants are the only signature of the function $f$ in the linearized equations. In particular, observe that when $\Lambda=0$, \ie for Minkowski spacetime, all sign from the higher-curvature terms of order cubic or higher disappears from the linearized equations.

For arbitrary values of $\Lambda$, the full linearized equation \req{lini} in terms of $h_{\mu\nu}$ reads
\begin{equation}
\begin{aligned}
&\alpha\left[\bar\nabla_{(\mu|}\bar\nabla_{\sigma} h^{\sigma}_{|\nu)}-\frac{1}{2}\bar\nabla_{\nu}\bar\nabla_{\mu}h-\frac{1}{2}\bar\Box h_{\mu\nu}+\Lambda h_{\mu\nu}-\Lambda h \bar g_{\mu\nu}\right]+\\
&+\left[\bar g_{\mu\nu}\left(\Lambda \beta-\frac{\alpha}{2}\right)+\frac{\beta}{D-1}\left(\bar g_{\mu\nu}\bar \Box-\bar\nabla_{\mu}\bar\nabla_{\nu}\right)\right]\left[\bar\nabla_{\alpha}\bar\nabla_{\beta}h^{\alpha\beta}-\bar\Box h-\Lambda (D-1) h\right]=0\, ,
\end{aligned}
\label{linearfull}
\end{equation} 
 and its trace is given by
 \begin{equation}\label{lineartr}
 \begin{aligned}
 &\left[ D\Lambda \beta-\alpha(D/2-1)\right]\left[ \bar\nabla_{\mu} \bar\nabla_{\nu}h^{\mu\nu}- \bar\Box h-\Lambda(D-1) h\right]\\
 &-\Lambda\beta (D-1)  \bar\Box h+\beta \left[ \bar\Box \bar\nabla_{\mu} \bar\nabla_{\nu}h^{\mu\nu}- \bar\Box^2 h\right]=0.
 \end{aligned}
 \end{equation}
 The two equations above are not particularly illuminating. However, it is already noticeable the absence of terms of the form $\bar\Box^2 h_{\mu\nu}$ in \req{linearfull}. Such terms indicate the presence of ghost-like massive spin-2 fields and its absence is a nice feature of this class of theories. In order to make this statement more clear, we can exploit the `gauge' symmetry of the linearized equations under transformations of the form $\delta h_{\mu\nu}=\bar \nabla_{\mu} \xi_{\nu}+\bar \nabla_{\nu} \xi_{\mu}$. In particular, 
 we choose the following (transverse) gauge 
 \begin{equation}\label{pollas}
  \bar \nabla_{\mu}h^{\mu\nu}=\bar\nabla^{\nu} h\, .
 \end{equation}
  In this gauge, the linearized equation \req{linearfull} and its trace \req{lineartr} become
 \begin{align}\label{lineargauge}
& \alpha\left[\frac{1}{2}\bar\nabla_{\mu}\bar\nabla_{\nu} h-\frac{1}{2}\bar\Box h_{\mu\nu}+\Lambda h_{\mu\nu}\right]+\Lambda  \left[\alpha\frac{(D-3)}{2}-(D-1)\Lambda \beta\right]\bar g_{\mu\nu} h\\ \notag &-\Lambda \beta\left[\bar g_{\mu\nu}\bar \Box-\bar\nabla_{\mu}\bar\nabla_{\nu}\right]h=0\, .\\ \label{traceeq} &
\Lambda\left[\beta\bar\Box+D\Lambda \beta-\frac{\alpha(D-2)}{2}\right]h=0.
 \end{align}
These expressions reduce to those obtained in \cite{Lu:2011zk,Smolic:2013gz} and \cite{Bueno2} in the particular cases of $R^2$ and $f(R)$ gravities respectively. Observe that all quartic derivatives have disappeared from these equations. Let us now split $h_{\mu\nu}$ into its trace and traceless components as
\begin{equation}
 h_{\mu\nu}=\hat h_{\mu\nu}+\frac{1}{D}\bar g_{\mu\nu} h.
\label{traceless2}
\end{equation}
If we rewrite (\ref{lineargauge}) using this new tensor we are left with the following inhomogeneous equation
\begin{equation}
-\frac{\alpha}{2}\left[\bar\Box \hat h_{\mu\nu}-2\Lambda\hat h_{\mu\nu}\right]+\left[\frac{\alpha}{2}+\Lambda \beta\right]\left[\bar\nabla_{\mu}\bar\nabla_{\nu}h-\frac{\bar g_{\mu\nu}}{D}\bar\Box h\right]=0.
\label{tracelesseq}
\end{equation}
This is still not completely satisfactory, as it contains terms involving the trace. We can however find an homogeneous equation by defining a new traceless tensor\footnote{We follow the procedure presented in \cite{Smolic:2013gz}.}
\begin{equation}
t_{\mu\nu}=\hat h_{\mu\nu}-\frac{2\beta}{(D-2)\alpha}\left[\bar\nabla_{\mu}\bar\nabla_{\nu}h-\frac{\bar g_{\mu\nu}}{D}\bar\Box h\right].
\label{temunu}
\end{equation}
Indeed, by using (\ref{tracelesseq}) and \req{traceeq}, one finds
that $t_{\mu\nu}$ satisfies the equation
\begin{equation}\label{fulfin}
-\frac{\alpha}{2}\left[\bar\Box t_{\mu\nu}-2\Lambda t_{\mu\nu}\right]=0.
\end{equation}
This is the equation for a traceless and massless spin-2 field. Hence, in $f($Lovelock$)$ theories, $t_{\mu\nu}$ is the tensor that represents the usual graviton. The other physical propagating degree of freedom is of course $h$. Indeed, \req{traceeq} is the equation of a scalar field of mass
\begin{equation}
M^2=\frac{(D-2)\alpha}{2\beta}-D\Lambda\, .
\end{equation}
This is of course provided $\beta\neq 0$. In such a case, when the background is AdS$_D$, the holographic dictionary \cite{Maldacena,Gubser,Witten} tells us that $h$ is dual to a scalar operator $\mathcal{O}_{\Delta}$ in the $(D-1)$-dimensional boundary CFT with scaling dimension
\begin{equation}
\Delta=\frac{(D-1)}{2}+\sqrt{\frac{(D+1)^2}{4}+\frac{\alpha (D-2)L^2}{2\beta}}\, ,
\end{equation}
where we wrote $\Lambda=-1/L^2$. When $\beta/(\alpha L^2)$ is small and positive, $\mathcal{O}_{\Delta}$  is a highly-irrelevant positive-norm operator with $\Delta\simeq \sqrt{\alpha (D-2)L^2/(2\beta) }$. On the other hand, if  $\beta/(\alpha L^2)$ is small and negative, $\Delta$ becomes imaginary, and $h$ is a ghost-like field with tachyonic mass exceeding the Breitenlohner-Freedman (BF) bound \cite{Breitenlohner:1982bm,Breitenlohner:1982jf}. In that case, our $f($Lovelock$)$ theory would automatically be unstable if we interpreted it as a complete description rather than as an effective low energy theory.

\subsection{Theories without dynamical scalar}\label{nosca}
Something interesting happens when $\beta=0$. In that case, just like for Einstein, Lovelock or quasi-topological gravities, the scalar mode is absent, and \req{traceeq} just tells us that the transverse gauge condition \req{pollas} imposes the trace to vanish, \ie $h=0$. In those cases, the only physical field is the massless graviton $t_{\mu\nu}$. 

In fact, when $\beta=0$, the full equations of motion become second order for any gauge. Indeed, in that case \req{lini}
 becomes
\begin{equation}\label{lini2}
\alpha \,G^L_{\mu\nu}=8\pi G\,  T_{\mu\nu}\, ,
\end{equation}
where we have included the stress tensor of some additional matter fields in the right-hand side in order to stress that the overall factor $\alpha$ is non-trivial, as it determines the normalization of Newton's constant: $G_{\rm eff}=G/\alpha$. Hence, we observe that for these theories, the linearized equations are exactly the same as for Einstein gravity but with an effective Newton constant controlled by $\alpha$.

Now, interestingly enough, $\beta=0$ does not necessarily imply $\lambda=0$, \ie there are non-trivial $f($Lovelock$)$ theories satisfying \req{lini2} and for which the only propagating degree of freedom is therefore the ususal graviton. These are characterized by the conditions
\begin{eqnarray}
\label{aaaaa}
\sum_{n,m=1}^{\lfloor D/2 \rfloor}\partial_n\partial_m f\left(\bar{\mathcal{L}}\right) \frac{n\,m(D-2)!(D-1)!}{(D-2n)!(D-2m)!}\,\Lambda^{n+m-2}=0\, , \quad \partial_n\partial_m f\left(\bar{\mathcal{L}}\right)\neq 0\, ,
\end{eqnarray}
for some $n$, $m$. Of course, these are satisfied by infinitely many classes of $f($Lovelock$)$ gravities. For example, theories of the form

\begin{equation}
S=\frac{1}{16\pi G}\int_{\mathcal{M}}d^Dx\sqrt{|g|}\left[-2\Lambda_0+R+\lambda  \left(R^v \mathcal{L}_2^w-\gamma R^{2w+v}\right)\right]\, ,
\label{parafernalia}
\end{equation}
where
\begin{equation}
\gamma=\frac{v^2+4(w-1)w+v(4w-1)}{(v+2w)(v+2w-1)}\frac{(D-2)^w(D-3)^w}{D^w(D-1)^w}\, ,
\end{equation}
for some $v,w\geq0$,
satisfy the requirements. If we choose $v=w=1$, $D=4$ in \req{parafernalia}, we find

\begin{equation}
S=\frac{1}{16\pi G}\int_{\mathcal{M}}d^4x\sqrt{|g|}\left[-2\Lambda_0+R+\lambda  \left(R \mathcal{L}_2-\frac{1}{9} R^{3}\right)\right]\, .
\label{parafernalia3}
\end{equation}
This is an example of a non-trivial four-dimensional cubic-order theory of gravity with second-order linearized equations of motion and which therefore only propagates a single massless graviton around m.s.b. The action \req{parafernalia3} is somewhat reminiscent of critical gravity \cite{Lu:2011zk}, a four-dimensional quadratic-theory for which the scalar degree of freedom is also absent, and the extra spin-2 field is massless. We stress again that all $f($Lovelock$)$ gravities are free of such spin-2 fields, so all theories satisfying \req{aaaaa}, like \req{parafernalia3}, have in this sense a better behavior than critical gravity: the scalar is also absent and there is no need to set the mass of the extra graviton to zero because there is no extra graviton at all either.  Also, recall that quasi-topological gravity exists only for $D\geq 5$, and that all Lovelock theories but Einstein gravity are trivial --- or topological in the case of Gauss-Bonnet --- in four dimensions. This makes \req{parafernalia3} --- and the rest of $D=4$ theories satisfying \req{aaaaa} --- particularly interesting and worth further study in our opinion.


\subsection{Comments on unitarity}
The propagator of $h_{\mu\nu}$ in any perturbatively unitary higher-curvature gravity around a m.s.b. is equal to the propagator of a quadratic theory of the form 
\begin{equation}
S=\frac{1}{16\pi G}\int_{\mathcal{M}}d^Dx\sqrt{|g|}\left[-2\Lambda_0+R+c_1 R^2+c_2 \mathcal{L}_2\right]\, .
\label{parafernalia4}
\end{equation}
In particular, the parameters of the corresponding higher-curvature theory are related to $G$, $\Lambda_0$, $c_1$ and $c_2$ above. These parameters are in turn constrained to satisfy different inequalities in order for the theory to be unitary --- essentially these come from imposing that the effective Newton constant is positive and that the mass of the scalar mode is positive for dS$_D$ and greater than the BF bound for AdS$_D$. We refer to \cite{Sisman:2011gz} for details --- see also \cite{Alvarez-Gaume:2015rwa}. Observe that in all cases, the massive spin-2 graviton is absent, given that the $R^2$ and GB terms do not introduce it. This is unsurprising, given that such field is generically a ghost and spoils unitarity. Hence, in general, the massless graviton and the scalar are the only allowed degrees of freedom in a unitary theory. Whenever $c_1 = 0$, the scalar will also be absent, leaving us with the usual massless graviton, and nothing else. The theories considered in the previous subsection belong to this class. For these, the linearized equations are second-order in any gauge, as we already stressed.

 The problem of classifying or identifying in full generality which higher-curvature theories share propagator with any of the curvature-square gravities in \req{parafernalia4} --- \ie which of them are unitary on m.s.b. provided the appropriate constraints are satisfied --- is non-trivial in general. In \cite{Sisman:2011gz}, the authors carried out this classification for the most general gravity theory constructed from contractions of the metric and the Riemann tensor at cubic order in curvature and in general dimensions. Using their results, it is not difficult to check that the three cubic $f($Lovelock$)$ terms constructed as products of ED --- namely $R^3$, $R\mathcal{L}_2$ and $\mathcal{L}_3$ --- belong to this class of theories. More generally, we expect \emph{all} $f($Lovelock$)$ theories to be perturbatively unitary around m.s.b. as long as the appropriate constraints on the couplings are satisfied. We leave a thorough exploration of this issue for future work.


\section{Holographic constraints on the coupling values} 
\labell{ee}
Holography \cite{Maldacena,Gubser,Witten} has become one of the main motivations for the study of higher-derivative gravities. As mentioned in the introduction, these theories have been used to characterize various properties of general CFTs in various dimensions --- see \eg \cite{Myers:2010xs,Myers:2010tj,Brigante:2007nu,Bueno1,Bueno2,Mezei:2014zla}. In order for a higher-derivative theory to admit a physically sensible dual description in the holographic context, it must satisfy certain requisites. Such requisites --- which have been previously considered many times in the past, \eg \cite{Buchel:2009sk,Camanho:2013pda,Camanho:2014apa,Banerjee:2014oaa} --- generically translate into constraints on the allowed values of the gravitational couplings. The most obvious example is the requirement that the theory admits at least one AdS vacuum --- otherwise one cannot even talk about any `dual theory'! Other considerations which in general lead to constraints on the gravity couplings consist of asking the dual theory to respect causality, unitarity, or certain quantum information inequalities.

 In this section, we will find constraints on the allowed values of the gravitational couplings of a particular class of $f($Lovelock$)$ theories. The first set of constraints will come from imposing AdS$_D$ to be a solution of the corresponding theory. For the second, we will restrict ourselves to $D=5$, and we will use holographic entanglement entropy (HEE) --- see subsection \ref{hee} for details.


A particularly relevant subclass of $f($Lovelock$)$ theories which appears several times throughout this paper is the one consisting of linear combinations of arbitrary products of ED. In general dimensions, we have the following possibilities at each other in curvature
\begin{eqnarray}\notag
 R ,\\ \notag
 R^2,\, \mathcal{L}_2 ,\\\notag
 R^3,\, R \mathcal{L}_2,\,  \mathcal{L}_3,\\\notag
 R^4,\, R^2  \mathcal{L}_2,\,  R\mathcal{L}_3,\,  \mathcal{L}_2^2,\,  \mathcal{L}_4,\\ \notag 
R^5,\, R^3 \mathcal{L}_2,\,  R^2\mathcal{L}_3,\, R\mathcal{L}_2^2,\,  R\mathcal{L}_4, \,\mathcal{L}_2\mathcal{L}_3,\, \mathcal{L}_{5}, \\ \dots
\end{eqnarray}
There are $1,2,3,5,7,11,15,22,30,42,\dots$ of these. Note that at $p$-th order in curvature, the number of terms is given by the so-called \textit{Partition Function} $P(p)$, which counts the number of ways in which the integer $p$ can be written as a sum of positive integers.\footnote{For example, for $p=4$ we have $4=1+1+1+1=3+1= 2+2= 2+1+1=4$, so $P(4)=5$. $P(p)$ also coincides with the number of conjugacy classes of the permutation group of order $p$.} In four and five dimensions, the most general Lagrangian density of this kind corresponds to a linear combination of terms of the form `$R^v\mathcal{L}_2^w$', where $v\,,w\in \mathbb{N}$. It will be for this last class of theories that we will construct the constraints.

\subsection{AdS$_{D}$ embedding}

As we have just anticipated, let us for now focus on the following $D$-dimensional subclass of  $f($Lovelock$)$ theories

\begin{equation}\label{rl2}
S_{v,w}=\frac{1}{16\pi G}\int_{\mathcal{M}} d^{D}x\,\sqrt{|g|}\left[ R+\frac{(D-1)(D-2)}{%
\tilde{L}^{2}}+\tilde{L}^{(2v+4w-2)}\lambda _{v,w}R^{v}(\mathcal{L}_{2})^{w}%
\right] ,\,
\end{equation}
where we have chosen the cosmological constant to be negative, $\Lambda_0=-(D-1)(D-2)/(2\tilde{L}^2)$, and where $\lambda_{v,w}$ is a dimensionless coupling. 
When $\lambda_{v,w}=0$, the embedding equation for AdS$_D$, whose metric in Poincar\'e coordinates reads
\begin{equation}
ds^{2}=\frac{L^{2}}{z^{2}}\left[ -dt^{2}+dz^{2}+d\vec{x}_{(D-2)}^{2}\right] \, ,
\label{Ads}
\end{equation}
simply imposes that the scale $\tilde{L}$ in the action is equal to the AdS radius $L$. Of course, as we include additional higher order terms, this is no longer true, and the relation between both scales depends on the new gravitational
couplings. For general $f($Lovelock$)$ theories, the corresponding embedding equation is \req{lambdaconst}. Applying it to the particular case of \req{rl2}, it becomes
%
%
%
\begin{equation}
1-f_{\infty }-C_{v,w}f_{\infty }^{v+2w}\lambda _{v,w}=0\, ,  \label{emb}
\end{equation}
where 
\begin{equation}
C_{v,w}=(-1)^{v-1}(D-1)^{w+v-1}D^{w+v-1}(D-2) ^{w-1}(D-3) ^{w} (D-2 ( v+2w))\, ,
\end{equation}
and where we have defined $f_{\infty }= \tilde{L}^{2}/L^{2}$.
It is not possible to solve the above equation for $f_{\infty}$ in full
generality. However, it suffices for our purposes to obtain the set of
values of $\lambda_{v,w}$ for which the above equation is satisfied in a
physically sensible way. In particular, we need to require $f_{\infty}$ to
be positive and tend to one as $\lambda_{v,w}\rightarrow0$. Using \req{emb}, we can write $\lambda _{v,w}$ as a function of $f_{\infty }$, \ie
\begin{equation}
\lambda _{v,w}(f_{\infty })=\frac{1-f_{\infty }}{C_{v,w}\text{ }%
f_{\infty }^{v+2w}}\, .
\end{equation}
Now, when $C_{v,w}>0,(C_{v,w}<0)$ the function $\lambda_{v,w}(f_{\infty })$ has a global
minimum (maximum) at
\begin{equation}
f^{*}_{\infty }=\frac{(2w+v)}{(2w+v-1)}\, .
\end{equation}
This directly implies the following constraint on the coupling constant $\lambda _{v,w}$,
\begin{eqnarray}\label{constraat}
\lambda _{v,w}\geq \lambda _{v,w}(f^{*}_{\infty })\, , \quad \text{when} \quad C_{v,w}>0\, , \\ \notag
\lambda _{v,w}\leq \lambda _{v,w}(f^{*}_{\infty })\, , \quad \text{when} \quad C_{v,w}< 0\, , 
\end{eqnarray}
and where
\begin{equation}
\lambda _{v,w}(f^{*}_{\infty })=-\frac{(2w+v-1)^{2w+v-1}}{C_{v,w}\, (2w+v)^{2w+v}}\, .
\end{equation}
For example, if we choose $v=0$, $w=1$, and define $\lambda_{0,1}=\lambda_{\rm \ssc GB}/((D-3)(D-4))$ --- as it is customary --- \req{rl2} becomes the usual GB theory, and \req{constraat} is nothing but the well-known constraint $\lambda_{\rm \ssc GB}\leq 1/4$, \cite{Boulware:1985wk,Buchel:2009sk,Myers:2010ru}. Another familiar example corresponds to $v=2$ and $w=0$, which is nothing but $R^2$ gravity. The constraint reads in that case $\lambda_{2,0}\leq(D-2)/(2(D-1)(D-4))$. Interestingly, this does not impose any constraint in four dimensions, which is a consequence of the fact that the couplings of general quadratic gravities do not enter into the embedding equation of AdS$_4$,\footnote{In other words, AdS$_4$ is a solution of general curvature-squared gravities --- and $R^2$ in particular --- as long as $L=\tilde{L}$, just like for Einstein gravity. We come back to this point in section \ref{sec:solutions}.}  see \eg \cite{Smolic:2013gz}.
Actually, we observe that this phenomenon occurs whenever
\begin{equation}\label{magic}
D=2(v+2w)\, .
\end{equation}
This means, for example, that AdS$_6$ is a solution of $R^3$ and $R\mathcal{L}_2$ gravities for arbitrary values of the corresponding gravitational couplings. 




%

\subsection{Holographic entanglement entropy}\label{hee}
In this subsection we will use holographic entanglement entropy (HEE) to find additional constraints on $\lambda_{v,w}$ for five-dimensional theories. Before explaining our procedure, let us start with some essentials about entanglement entropy in the holographic context.

Consider a bipartition of the Hilbert space of some quantum system, $\mathscr{H}=\mathscr{H}_{A}\otimes \mathscr{H}_{B}$, and some state $\rho$. The entanglement entropy (EE) is defined with respect to the reduced state corresponding to one of the partitions, say $A$, obtained by tracing out the degrees of freedom in $B$, $\rho_A=\tr_B \rho $. In particular, the EE is defined as the Von Neumann entropy of $\rho_A$, \ie $S_{\rm EE}(\rho_A)=-\tr (\rho_A \log \rho_A)$. In the following, we will restrict ourselves to spatial bipartitions, meaning that $A$ will always be a physical spatial region at a fixed time slice, and $B$ its complement.

In the context of holography, EE is computed using the Ryu-Takayanagi (RT) prescription \cite{Ryu:2006ef,Ryu:2006bv}. According to this, given an asymptotically AdS$_D$ spacetime dual to some state in the boundary theory, the HEE for a region $A$ in the corresponding CFT is obtained by extremizing the area functional of codimension-2 bulk surfaces $m$ which are homologous to $A$ in the boundary (and, in particular, $\partial m=\partial A$). More precisely,
\begin{equation}\label{rt}
S_{\rm RT}(A)=\underset{m\sim A}{\text{ext}}\left[\frac{\mathcal{A}(m)}{4G}\right]\, ,\quad \mathcal{A}(m)=\int_{m}d^{D-2}x\, \sqrt{h_m}\, ,
\end{equation}
where $G$ is the Newton constant and $h_m$ is the determinant of the metric induced on $m$. Naturally, this prescription is only valid for theories dual to Einstein gravity in the bulk. In particular, when higher-derivative terms are introduced in the bulk theory, the area functional in \req{rt} must be modified to something like 
\begin{equation}\label{rt2}
S(A)=\underset{m\sim A}{\text{ext}} S_{\rm grav}(m)\, ,
\end{equation}
where $S_{\rm grav}(m)$ is a new bulk functional which depends on the particular higher-derivative theory, and which reduces to \req{rt} for Einstein gravity. Much effort has been put into trying to identify the explicit form of $S_{\rm grav}(m)$ for different higher-derivative bulk theories, with remarkable success --- see \cite{Jacobson:1993xs,deBoer:2011wk,Hung:2011xb,Dong:2013qoa,Camps:2013zua,Bhattacharyya:2013jma,Fursaev:2013fta,Myers:2013lva,Lewkowycz:2013nqa,Dong:2015zba,Wall:2015raa} for a non-exhaustive list of references. In particular, a new functional consisting of a Wald-like term \cite{Wald:1993nt,Jacobson:1993vj,Iyer:1994ys} plus corrections involving extrinsic curvatures has been proposed to hold for general higher-derivative gravities \cite{Dong:2013qoa}. While such proposal passes various consistency checks, certain subtleties arise \cite{Bhattacharyya:2014yga,Miao:2014nxa,Astaneh:2014wxg,Huang:2015zua} when the theory is not Einstein, curvature-squared or Lovelock, which make it unclear how to use this prescription in general.\footnote{We thank Rong-Xin Miao for useful comments about this point.}

As explained in the introduction, the authors of \cite{Sarkar:2013swa} proposed \req{SW1} to be the right formula for the gravitational entropy in $f($Lovelock$)$ theories.
In that paper, the authors were able to show that this functional satisfies an increase theorem for linearized perturbations of Killing horizons. Besides, $S_{\rm \ssc SW}$ reduces to the well-known JM functional for Lovelock gravities \cite{Jacobson:1993xs} which, as already mentioned, gives rise to the right universal terms when used to compute HEE for these theories.  In \cite{Bueno1,Bueno2}, these two facts were interpreted as evidence that $S_{\rm \ssc SW}$ is in fact the right HEE functional for $f($Lovelock$)$ theories. The results found in those papers strongly support this claim.

Our plan is to use \req{SW1} to find new constraints on the coupling values $\lambda_{v,w}$. The idea \cite{Banerjee:2014oaa} is to consider simple entangling regions for which the surface $m$ can be parametrized as some function $g(z)$ of the holographic coordinate. While extremizing $S_{\rm \ssc SW}$ --- \ie finding the explicit form of $g(z)$ --- is an impracticable task in general, we do know that $m$ must close off smoothly at some bulk point $z=z_{h}$. Hence, we can assume that $g(z)$ admits a series expansion around $z_{h}$,
\begin{equation}
g(z)=\sum_{i=0}^{\infty }c_{i}(z_{h}-z)^{\alpha +i} \, . \label{smooth}
\end{equation}
Besides, we need to impose that $g(z_h)=0$ and $g^{\prime }( z_h)= -\infty $, since the tangent to the
surface will be perpendicular to the $z$ direction at that point. These conditions imply the constraints $0<\alpha <1$
and $c_{0}>0$, which we will use to find bounds on $\lambda _{v,w}$.

Evaluating \req{SW1} for our $f($Lovelock$)$ theory \req{rl2}, one finds
\begin{equation}
S_{v,w}=\frac{1}{4G}\int_{m}d^{3}x\sqrt{h_m}\,\left[ A+BL^{2}\,\mathcal{R}%
_{m}\right] \,,  \label{entro}
\end{equation}
where we defined the constants\footnote{Note that these are related to the constant $\alpha$ defined in \req{alpha} through: $\alpha=A-2B$.}
\begin{equation}\label{caracoles}
A=1+\lambda _{v,w}\,(-1)^{v-1}2^{2v+3w-2}5^{v+w-1}3^{w}v\,f_{\infty }^{(v+2w-1)}\,,\quad B=\frac{w(1-A)}{3v}\, ,
\end{equation}
and where $\mathcal{R}_m$ is the Ricci scalar associated to the induced metric on the holographic surface $m$. In particular, note that for $v=0$, $w=1$, $\lambda_{0,1}=\lambda_{\ssc \rm GB}/2$, \req{entro} reduces to the JM functional for GB gravity, \ie
\begin{equation}
S_{\rm \ssc JM}=\frac{1}{4G}\int_{m}d^{3}x\sqrt{h_m}\,\left[ 1+\lambda_{\ssc \rm GB}\tilde{L}^{2}\,\mathcal{R}%
_{m}\right] \,.  \label{entrojm}
\end{equation}
In \cite{Banerjee:2014oaa}, new constraints on $\lambda_{\ssc \rm GB}$ were obtained using this functional for some simple entangling regions following the procedure outlined above. Here we will generalize those results to arbitrary values of $v$ and $w$. A quick look at \req{entro} and \req{caracoles} shows that no bounds can be found using this technique for theories with $w=0$. The reason is that for those, the holographic extremal surface will be the same as in Einstein gravity, so it will not depend on the value of $\lambda_{v,0}$.

Let us start considering an entangling region consisting of a slab of width $l$ defined by $x_{1}\in [-l/2, \leq l/2]$, $x_{2}\in(-\infty,+\infty)$, $x_{3}\in(-\infty,+\infty)$. Now, using the obvious symmetry along the $x_{2,3}$ directions, we can parametrize the holographic surface $m$ as $t_{\rm E}=0$, $x_{1}=g(z)$. The induced metric on this surface reads
\begin{equation}
ds_{m}^{2}=\frac{L^{2}}{z^{2}}\left[ (1+\dot{g}^{2})dz^{2}+dx_{2}^{2}+dx_{3}^{2}\right] \,,
\end{equation}
where we used the notation $\dot{g}=dg(z)/dz$. The Euler-Lagrange equation for $g(z)$ obtained from \req{entro}, reads 
\begin{equation}\label{eomm1}
-3(A-2B)\dot{g}-6(A-B)\dot{g}^{3}-3A\dot{g}^{5}+(A-2B)\ddot{g}z+(A+4B)z\dot{g}^{2}\ddot{g}
=0\,.
\end{equation}
Inserting now the series expansion \req{smooth} in this equation, we find that the only value of $\alpha$ compatible with the smoothness requirements is $\alpha=1/2$. Using this, we can find the value of $c_0$ by imposing the coefficient of the lowest order term in \req{eomm1} to vanish. By doing so, we find
\begin{equation}
c_{0}=\sqrt{\frac{2(A+4B)z_h}{3A}}\, .
\end{equation}
This imposes $(A+4B)/A>0$ which, after some careful calculations, gives rise to the following constraint on $\lambda_{v,w}$ 
\begin{eqnarray}\label{constraa}
\lambda _{v,w}< \lambda^{\rm (s)} _{v,w}\, , \quad \text{when} \quad v \text{ even}\, , \\ \notag
\lambda _{v,w}> \lambda^{\rm (s)} _{v,w}\, , \quad \text{when} \quad v \text{ odd}\, , 
\end{eqnarray}
where
\begin{equation}\label{condee}
\lambda^{\rm (s)} _{v,w}=\frac{(-1)^v     (5 v+4
   w-5)^{v+2 w-1}}{v^{2 w+v}2^{2 v+3 w-2} 3^{v+3 w-1} 5^{v+w-1}}\, .
\end{equation}
and which is valid whenever $v\geq1$ and $w\geq1$. 
The GB case, $v=0$, $w=1$, is a bit special, and one finds
\begin{eqnarray}\label{constraa2}
\lambda _{\rm \ssc GB}=2\lambda _{0,1}>-\frac{5}{16}  \, ,
\end{eqnarray}
in agreement with the result of \cite{Banerjee:2014oaa}. For $v=0$ and $w>1$, no bounds on $\lambda_{0,w}$ are found.  Also, as a check of our procedure, we have verified that no bounds appear when $w=0$ --- indeed, $(A+4B)/A=1$ in that case.

Let us now consider a cylindrical entangling surface. We write the AdS$_{5}$ metric as
\begin{equation}
ds^{2}=\frac{L^{2}}{z^{2}}\left[ dt_{\ssc \rm E}^{2}+dz^{2}+d\rho ^{2}+\rho ^{2}d\theta
^{2}+dx_{3}^{2}\right] \,,
\end{equation}
and let the cylinder be defined as $t_{\ssc \rm E}=0$, $\rho \in \lbrack 0,R]$, $\theta \in \lbrack 0,2\pi )$, $x_{3}\in ( -\infty,+\infty)$. Again, the symmetry of the entangling surface allows us to parametrize the holographic surface as $t_{\ssc \rm E}=0$, $\rho =g(z)$. The pullback metric on such surface reads
\begin{equation}
ds_{m}^{2}=\frac{L^{2}}{z^{2}}\left[ (1+\dot{g}^{2})dz^{2}+g^{2}d\theta
^{2}+dx_{3}^{2}\right] \,.
\end{equation}
The corresponding Euler-Lagrange equation for $g(z)$ reads in this case
\begin{align}\notag
&-(1+\dot{g}^{2})\left[ 3g\dot{g}\left[ A(1+\dot{g}^{2})-2B\right] +z\left[
A-2B+(A+4B)\dot{g}^{2}\right] \right] \\
&+z\left[ 6Bz\dot{g}+g(A-2B+(A+4B)\dot{g%
}^{2})\right] \ddot{g}=0\,.
\end{align}
From this we find again $\alpha =1/2$, and using this result we obtain the only allowed value of $c_0$ to be
\begin{equation}
c_{0}= \sqrt{\frac{2z_{h}}{3A}\left[ (A+4B)\pm \sqrt{A^{2}+16B^{2}-10AB}%
\right] }\, ,
\end{equation}
so we are lead to impose $A^{2}+16B^{2}-10AB\geq 0$.
%
%
%
A careful analysis shows that the condition that follows from this inequality reads
\begin{eqnarray}\label{constraa3}
\lambda _{v,w}\leq \lambda^{\rm (c)} _{v,w}\, , \quad \text{when} \quad v \text{ even}\, , \\ \notag
\lambda _{v,w}\geq \lambda^{\rm (c)} _{v,w}\, , \quad \text{when} \quad v \text{ odd}\, , 
\end{eqnarray}
where
\begin{equation}
\lambda^{\rm (c)} _{v,w}=\frac{(-1)^{v}(-5+5v+12w)^{v+2w-1}}{
3^{w-1}2^{3w+2v-2}5^{w+v-1}(3v+8w)^{v+2w}}\,.
\end{equation}
As opposed to the slab --- for which the GB case was special --- this condition is the same for all values $v\geq0$ and $w\geq1$. In particular, one finds
\begin{eqnarray}\label{constraa2}
\lambda _{\rm \ssc GB}=2\lambda _{0,1}\leq \frac{7}{64}  \, ,
\end{eqnarray}
again in agreement with the bound found in \cite{Banerjee:2014oaa}. We have checked again that no bounds are found when $w=0$, as expected.


%
%
%
%
%
%
%
%
%
 %
%


In sum, for the family of theories \req{rl2} in five dimensions, we have found constraints from the AdS$_5$ embedding, and from imposing the holographic surface corresponding to a slab and a cylindrical entangling region to close off smoothly in the bulk. Combining all these constraints, we found the following bounds on $\lambda_{v,w}$, 
\begin{eqnarray}\label{constraa}
\lambda _{v,w}(f_{\infty}^*)\leq\lambda _{v,w}\leq \lambda^{\rm (c)} _{v,w}\, , \quad \text{when} \quad v \text{ even}\, , \\ \notag
\lambda _{v,w}(f_{\infty}^*)\geq\lambda _{v,w}\geq \lambda^{\rm (c)} _{v,w}\, , \quad \text{when} \quad v \text{ odd}\, , 
\end{eqnarray}
which are valid for $v\geq 1$, $w\geq 1$ and $v=0$, $w>1$. Note that these are quite strong constraints in general. For example, one finds $\lambda _{1,1}(f_{\infty}^*)=1/270\simeq 0.0037$, $\lambda^{\rm (c)} _{1,1}=-18/6655\simeq -0.0027$. Further, for larger values of $v$ and $w$, the quantities $\lambda _{v,w}(f_{\infty}^*)$ and $\lambda^{\rm (c)} _{v,w}$ become increasingly smaller. In the case of Gauss-Bonnet, the bounds read instead $-5/16\leq\lambda _{\rm \ssc GB}\leq 7/64$. Finally, recall that when $w=0$, we only have the bound from the AdS$_D$ embedding, \ie the one given in \req{constraat}.



\section{Black hole solutions} 
\label{sec:solutions}

In this section we construct analytic solutions of several $f($Lovelock$)$ theories in various dimensions. 

\subsection{Pure $f($Lovelock$)$}
Let us start considering a $f($Lovelock$)$  theory consisting of a function of a single ED, $\mathcal{L}_n$, \ie
\begin{equation}\label{pure}
S=\frac{1}{16\pi G}\int_{\mathcal{M}}d^Dx\sqrt{|g|}\,f(\mathcal{L}_n)\, .
\end{equation}
Of course, for $n=1$, this reduces to $f(R)$ gravity, whose constant $R$ solutions were first studied in \cite{Maroto}. Here we will generalize some of their results by constructing constant-$\mathcal{L}_n$ solutions to \req{pure} for arbitrary values of $n$.
The field equations of the theory read
\begin{equation}\label{pepe}
f'(\mathcal{L}_n)\mathcal{E}_{\mu\nu}^{(n)}+\frac{1}{2}g_{\mu\nu}\left[\mathcal{L}_{n}f'(\mathcal{L}_n)-f(\mathcal{L}_n)\right]-2P^{(n)}_{\alpha\nu\lambda\mu}\nabla^{\alpha}\nabla^{\lambda}f'\left(\mathcal{L}_n\right)=0\, ,
\end{equation}
where the tensors $\mathcal{E}_{\mu\nu}^{(n)}$ and $P^{(n)}_{\alpha\nu\lambda\mu}$ were defined in \req{love}. In particular, note that the equations of motion of a pure Lovelock theory consisting of a single ED of order $n$ plus a cosmological constant term, \ie 
\begin{equation}\label{puri}
f(\mathcal{L}_n)=-2\Lambda_0+\Lambda_0^{1-n}\mathcal{L}_n\,, 
\end{equation}
satisfy 
\begin{equation}\label{puradroga}
\mathcal{E}_{\mu\nu}^{(n)}=-\Lambda_0^{n}g_{\mu\nu}\, , \quad \text{which implies} \quad \mathcal{L}_n=\frac{2D}{(D-2n)}\Lambda_0^n\,.
\end{equation}
In other words, all solutions of \req{puri} have a constant $\mathcal{L}_n$ proportional to the $n$-th power of the cosmological constant, just like all solutions of general relativity in the absence of matter have a constant Ricci scalar proportional to $\Lambda_0$.

Now, let us see under what conditions spacetimes of constant $\mathcal{L}_n$ solve the general $f(\mathcal{L}_n)$ equations \req{pepe}. If we assume $\mathcal{L}_n$ to be constant, the term with the covariant derivatives vanishes, and we are left with
\begin{equation}
f'(\mathcal{L}_n)\mathcal{E}_{\mu\nu}^{(n)}+\frac{1}{2}g_{\mu\nu}\left[\mathcal{L}_{n}f'(\mathcal{L}_n)-f(\mathcal{L}_n)\right]=0\, ,
\label{oneEDeq}
\end{equation}
whose trace reads
\begin{equation}
nf'(\mathcal{L}_n)\mathcal{L}_{n}-\frac{D}{2}f(\mathcal{L}_n)=0\, .
\label{algebraicLn}
\end{equation}
Given a particular $f$, this is an algebraic equation which solutions of \req{oneEDeq} are forced to satisfy. In particular, assuming it admits a solution, \req{algebraicLn} fixes $\mathcal{L}_n$ to some constant value which we will denote $\mathcal{L}^{0}_n$. Now we have two possibilities, depending on whether the derivative of $f$ vanishes when evaluated on $\mathcal{L}^{0}_n$. If $f'(\mathcal{L}^{0}_n)\neq 0$, we recover the pure Lovelock field equations, while if $f'(\mathcal{L}^{0}_n)= 0$, we do not need to impose any additional condition, because, in that case, all configurations satisfying $\mathcal{L}_n=\mathcal{L}^{0}_n$ are already extremal points of the action. Let us explain both cases in more detail.

Assume first that $f'(\mathcal{L}^{0}_n)\neq 0$. In that case, we can rewrite \req{oneEDeq} as
\begin{equation}
\mathcal{E}^{(n)}_{\mu\nu}=-\Lambda^n_{0,\rm eff}\,g_{\mu\nu}\, , \quad \text{where}\quad  \Lambda^n_{0,\rm eff}=\frac{(D-2n)}{2D}\mathcal{L}^{0}_n\, ,
\label{LLequiv2}
\end{equation}
which is nothing but the pure Lovelock equation of motion \req{puradroga} with an effective cosmological constant $ \Lambda_{0,\rm eff}$ determined by the solution of \req{algebraicLn}. Hence, any solution of pure Lovelock plus cosmological constant, is also a constant-$\mathcal{L}_n$ solution of \req{pure} provided $f'(\mathcal{L}^{0}_n)\neq 0$. This allows, in particular, to embed all Einstein gravity plus cosmological constant solutions in $f(R)$ whenever $f^{\prime}(R^0)\neq 0$, as explained in \cite{Maroto}.

Static black hole solutions of pure Lovelock gravities have been previously considered several times --- see \eg \cite{PureLove,PureLove2,Banados:1993ur,Cai:1998vy,Wheeler:1985qd}. In particular, for $D>2n$, a theory of the form \req{puri} admits the following interesting generalization of the Schwarzschild(-AdS/dS) black hole solution \cite{PureLove}
\begin{equation}
ds^2=-g(r)dt^2+\frac{1}{h(r)}dr^2+r^2d\Omega_{(D-2)}^2\, ,
\label{2functmetric}
\end{equation}
where
\begin{equation}
g(r)=h(r)=1\mp  r^2\left[\frac{a}{r^{D-1}}+\Lambda_0^n\frac{2 (D-2n-1)!}{(D-1)!}\right]^{1/n}\, .
\label{solution1}
\end{equation}
In this expression, $a$ is an integration constant which can be related to the solution's mass, and the $+$ sign in front of the term in brackets is allowed only when $n$ is even.
According to our analysis above, this is also a solution of \req{pure} for theories satisfying $f'(\mathcal{L}^{0}_n)\neq 0$. More precisely, using \req{LLequiv2} we find that the solution to \req{pure} can be written as \req{2functmetric} with 
\begin{equation}
g(r)=h(r)=1\mp r^2\left[\frac{a}{r^{D-1}}+\mathcal{L}^{0}_n\frac{(D-2n)!}{D!}\right]^{1/n}\, ,
\label{solution16}
\end{equation}
where again $\mathcal{L}^{0}_n$ is a solution to \req{algebraicLn}. For $n=1$, this reduces to the well-known $f(R)$ Schwarzschild(-AdS/dS) black hole \cite{Maroto}
\begin{equation}
g(r)=h(r)=1-\frac{a}{r^{D-3}}-\frac{R^0}{D(D-1)} r^2\, .
\label{solution123}
\end{equation}
For general values of $n$, \req{2functmetric} and \req{solution16} describe a $f($Lovelock$)$ generalization of the Schwarzschild(-AdS/dS) solution. 

Note that if the dimension is the critical one, $D=2n$, (\ref{LLequiv2}) is trivially satisfied because $\mathcal{E}_{\mu\nu}^{(D/2)}= 0$ identically. In that case, any solution of \req{algebraicLn} --- \ie any constant-$\mathcal{L}_n$ spacetime --- is a solution of \req{pure}. For example, in $D=4$, $f(\mathcal{L}_2)$ always allows for a solution with $\mathcal{L}_2=\mathcal{L}^{0}_2$. 

Before we turn to the $f'(\mathcal{L}^{0}_n)= 0$ case, let us make a further observation. Assuming $D>2n$, let us consider a theory of the form
\begin{equation}\label{puritano}
f(\mathcal{L}_n)=-2\Lambda_0+\Lambda_0^{1-n}\mathcal{L}_n+\alpha\Lambda_0^{1-\frac{D}{2}}\mathcal{L}_n^{\frac{D}{2n}}\,, 
\end{equation}
for some dimensionless constant $\alpha$. Interestingly, all solutions of the $\alpha=0$ theory are also solutions of \req{puritano}. This can be easily seen by imposing the $\alpha=0$ equation of motion \req{puradroga} in the equations \req{pepe} and \req{algebraicLn} corresponding to \req{puritano}. By doing so, we observe that the terms proportional to $\alpha$ exactly cancel each other out. This explains, in particular, why all solutions of Einstein gravity plus cosmological constant are also solutions of such theory with an additional $R^2$ term in four dimensions \cite{Smolic:2013gz}. Hence, we observe that \req{2functmetric} with \req{solution1} is also a solution of \req{puritano}. The reason for this general behavior can be traced back to the fact that $\mathcal{L}_n^{D/2n}$ is scale-invariant, \ie it is preserved by a rescaling of the metric. Now, a rescaling of the metric changes the scale of the theory. Hence, such scale cannot depend on $\alpha$.
 
Let us now turn to the cases for which  $f'(\mathcal{L}^0_n)=0$. If this happens, \req{algebraicLn} imposes also $f(\mathcal{L}^0_n)=0$ and the equations of motion \req{oneEDeq} are automatically satisfied. This means that spacetimes of constant-$\mathcal{L}_n$ are solutions of \req{pure} when these two conditions are satisfied. Obviously, this happens because a configuration $\mathcal{L}^0_n$ satisfying 
\begin{equation}\label{pepino}
f'(\mathcal{L}^0_n)=f(\mathcal{L}^0_n)=0\, ,
\end{equation}
 is always an extremum of the action. The existence of this kind of configurations depends on the particular theory under consideration. 
The simplest example is probably $f(\mathcal{L}_n)=\mathcal{L}_n^2$, for which $\mathcal{L}^0_n=0$ is clearly an extremum of the action and therefore a solution. The lesson is that in order to find a solution for these theories, we only need to require $\mathcal{L}_n$ to be equal to the constant $\mathcal{L}^0_n$ for which \req{pepino} holds. Since this is a single scalar equation for the metric, the number of possible solutions is huge. In particular, for an ansatz of the form \req{2functmetric} with $h(r)=g(r)$, one gets
\begin{equation}
g(r)=1\mp r^2\left[\frac{a}{r^{D-1}}+\frac{b}{r^{D}}+\mathcal{L}^{0}_n\frac{(D-2n)!}{D!}\right]^{1/n},
\label{singularsolution}
\end{equation}
which has two integrations constants, $a$ and $b$, instead of one. For $n=1$, this reduces to
\begin{equation}
g(r)=1-\frac{a}{r^{D-3}}-\frac{b}{r^{D-2}}-\frac{R^0}{D(D-1)} r^2\, ,
\label{singularsolution1}
\end{equation}
as observed in \cite{Maroto}. Note that for $D=4$, this takes the familiar form of the Reissner-Nordstr\"om(-AdS/dS) solution, with $b$ playing the role of the charge squared. Observe that this fact is accidental, and occurs only in four dimensions. In fact, it can be easily seen that the $f($Lovelock$)$-Maxwell system
\begin{equation}
	S=\int_{\mathcal{M}}d^Dx\sqrt{|g|}\left[\frac{1}{16\pi G}f\left(\mathcal{L}_n\right)-\frac{1}{4}F_{\mu\nu}F^{\mu\nu}\right]\, ,
\end{equation}
admits the following generalization of the Reissner-Nordstr\"om(-AdS/dS) solution 
\begin{equation}\label{chori}
g(r)=h(r)=1\mp r^2\left[\frac{a}{r^{D-1}}-\frac{c}{r^{2(D-2)}}+\mathcal{L}^{0}_n\frac{(D-2n)!}{D!}\right]^{1/n},
\end{equation}
when $f'(\mathcal{L}^{0}_n)\neq 0$, where $\mathcal{L}^{0}_n$ is again a solution of \req{algebraicLn} and where $c$ is a constant related to the electric charge. Comparing \req{singularsolution} with \req{chori}, we see that only when $D=4$ the exponents of the terms proportional to $b$ and $c$ respectively are equal.

If one considers the general ansatz (\ref{2functmetric}) with two unknown functions, the system is underdetermined, since we only have one equation. For example, let us consider $f(\mathcal{L}_2)=\mathcal{L}_2^2$ in $D=4$. In this theory, a family of solutions is given by $\mathcal{L}_2=0$. Assuming the ansatz (\ref{2functmetric}), we get the following equation for $g$ and $h$
\begin{equation}
\frac{h(r)^{1/2}\left[h(r)-1\right]}{g(r)^{1/2}}\frac{dg(r)}{dr}=c_1
\end{equation}
where $c_1$ is an integration constant. Choosing one of the functions at will, one can find the other by solving the above equation. This approach was followed, \eg in \cite{Kehagias:2015ata} to construct solutions to pure $R^2$ gravity.\footnote{See also, \eg \cite{Nojiri:2001aj}, where certain aspects of AdS black holes in pure curvature-squared gravities where considered.}

\subsection{General $f($Lovelock$)$}
In the previous subsection, we restricted ourselves to the case of $f($Lovelock$)$ theories consisting of functions of a single ED. Let us now explore what the situation is when one considers a function which depends on all the non-vanishing ED's, $f(\mathcal{L}_1,...,\mathcal{L}_{\lfloor D/2 \rfloor})$. Just like we were able to embed all solutions of the pure Lovelock Lagrangian \req{puri} in $f(\mathcal{L}_n)$, we would like to embed solutions of the general Lovelock action \req{cases} in the general $f($Lovelock$)$ one \req{flov}.
A simple argument shows that, in general, $f($Lovelock$)$ theory does not contain all solutions of Lovelock. The reasoning goes as follows. Assume that all solutions of the Lovelock equations
\req{efl} are also solutions of $f($Lovelock$)$ theory (\ref{fLL1}). Then, the ED associated to these metrics satisfy the trace equation
\begin{equation}
\sum_{n=0}^{\lfloor D/2 \rfloor}\lambda_n\Lambda_0^{1-n}\left(n-\frac{D}{2}\right)\mathcal{L}_n=0\, ,
\label{onshellED}
\end{equation}
but now, assume that these metrics also solve \req{fLL1}. Then, \req{fLL1} should reduce to \req{efl} whenever \req{onshellED} is satisfied. This implies, in particular, that all the partial derivatives of $f$ evaluated on the solution must be constant, which of course is not true for arbitrary functions $f(\mathcal{L}_1,...,\mathcal{L}_{\lfloor D/2 \rfloor})$. Hence, we observe that, on general grounds, solutions of Lovelock gravity are not embeddable in $f($Lovelock$)$ unless $f$ is chosen in an appropriate way --- like we did in the previous section by making it depend on a single ED.

We claim that the most general $f($Lovelock$)$ theory whose solutions include all the Lovelock theory ones is given by a function of the form
\begin{equation}
f(\mathcal{L}_1,...,\mathcal{L}_{\lfloor D/2 \rfloor})=\alpha\sum_{n=0}^{\lfloor D/2 \rfloor}\lambda_n\Lambda^{1-n}_0\mathcal{L}_n+\left[\sum_{n=0}^{\lfloor D/2 \rfloor}\lambda_n\Lambda^{1-n}_0\left(n-\frac{D}{2}\right)\mathcal{L}_n\right]^2 \tilde{f}(\mathcal{L}_1,...,\mathcal{L}_{\lfloor D/2 \rfloor})\, ,
\end{equation} 
where $\alpha$ is a constant and $\tilde{f}$ is an arbitrary function such that its derivatives are non-singular when the squared quantity is zero. 

\subsection{A critical black hole in $f(R,\mathcal{L}_2)$}
As we have seen, finding black solutions to $f($Lovelock$)$ theories involving more than one ED seems to be a difficult challenge. An exception is, of course, the case in which $f$ is a linear combination of ED, corresponding to general Lovelock theories --- see \eg \cite{Boulware:1985wk,Wiltshire:1988uq,Cai:2001dz,Cai:2003gr,Garraffo:2008hu,Camanho:2011rj}. A possible simplification that has been often considered in the literature for other higher-derivative gravities --- see \eg \cite{Cai:2009ac, AyonBeato:2010tm}, consists of fixing some of the couplings of the theory to particular values which allow for solutions which would not exist otherwise. We will explore this approach here. In particular, let us consider the following 
$f(R,\mathcal{L}_2)$ theory --- whose general equations of motion are specified in appendix \ref{frl2} --- consisting of the standard Einstein-Hilbert action plus certain higher-order corrections 
\begin{equation}
\label{BPSaction}
S=\int_{\mathcal{M}} \frac{d^D x \sqrt{-g}}{16\pi G} \left[ R+\frac{(D-1)(D-2)}{\tilde L^2}+\alpha \tilde L^2 R^2+\beta \tilde L^2 \mathcal{L}_2+\gamma \tilde L^4 R \mathcal{L}_2+\delta \tilde L^6 \mathcal{L}_2^2 \right] \, ,
\end{equation}
where $\alpha$, $\beta$, $\gamma$ and $\delta$ are dimensionless constants, and where we have chosen the cosmological constant to be negative and determined by some length scale $\tilde{L}$.
If the coupling parameters satisfy
\begin{align} \label{nohaymasnada}
\alpha = \frac{1}{4(D-1)(D-2)} \, , \quad \beta = \frac{\lambda}{(D-2)(D-3)} \, , \quad \gamma= 2\alpha\beta \, , \quad \delta = \alpha\beta^2 \, ,
\end{align}
then \req{BPSaction} allows for the following solution
\begin{equation}
ds^2=-g(r) dt^2+\frac{1}{g(r)}dr^2+r^2 d\Omega^2_{(D-2)} \, , 
\end{equation}
with
\begin{equation}
\label{eq:BPS}
g(r)=1+\frac{r^2}{2\lambda \tilde L^2}\left[1-\sqrt{1-4\lambda\left(\frac{2D-4}{D}-\frac{c_1}{r^{D-1}}+\frac{c_2}{r^D}\right)} \right] \, ,
\end{equation}
where $c_1$ and $c_2$ are integration constants. This solution describes an asymptotically AdS$_D$ black hole as long as the constants are chosen so that $g(r)>0$ for all $r>r_h$ and $g(r_h)=0$ for some positive value of $r$.

The reason why the election of parameters in \req{nohaymasnada} allows for this solution is not very mysterious. In fact, \req{nohaymasnada} makes the Lagrangian in \req{BPSaction} become a perfect square,
\begin{equation}
\label{eq:BPS2}
f(R,\mathcal{L}_2)=\frac{(D-1)(D-2)}{\tilde L^2}\left[1+\frac{\tilde L^2 R}{2(D-1)(D-2)}+\frac{\lambda \tilde L^4 \mathcal{L}_2}{2(D-1)(D-2)^2(D-3)}\right]^2 \, ,
\end{equation}
which implies that any configuration satisfying $f(R,\mathcal{L}_2)=0$ also satisfies $\partial_1 f(R,\mathcal{L}_2)=\partial_2 f(R,\mathcal{L}_2)=0$ and is therefore a solution of the corresponding equations of motion.

When $\lambda=0$, all the higher-derivative terms in \req{BPSaction} but the $R^2$ one disappear and \req{eq:BPS} takes the form of \req{singularsolution1}, \ie
\begin{equation}
g(r)=1-\frac{a}{r^{D-3}}-\frac{b}{r^{D-2}}+\frac{r^2}{\tilde L^2} \, ,
\end{equation}
as observed in \cite{AyonBeato:2010tm}. In \cite{Cai:2009ac,AyonBeato:2010tm}, the thermodynamic properties of various black hole solutions of \req{eq:BPS2} with $\lambda=0$ were studied. As observed there, while some of the solutions correspond to regular black holes with finite horizons, amusingly enough, they always possess vanishing entropy and mass. The reason is that both the on-shell action --- including boundary terms --- and the Wald entropy involve factors of either $f(R)$ or $f'(R)$, which vanish for these configurations, as we have just seen. In the case $\lambda\neq 0$, the situation is exactly the same. In particular, the action, the boundary term we have proposed in \req{bizcocho} --- necessary to compute the on-shell action --- and the entropy functional of \cite{Sarkar:2013swa} --- see \req{SW1} --- vanish on-shell for configurations of this kind. The physical interpretation of these solutions is unclear to us.





\section{Final comments and perspectives}\label{disc}
In this paper we have developed several aspects of $f($Lovelock$)$ theories. A summary of our main findings was already provided in section \ref{mr}, so we will not repeat it here. However, let us comment on some additional directions which would be worth exploring in the future. 

Note that we have followed a metric approach to $f($Lovelock$)$ theories. However, higher-derivative gravities can in general be studied using other methods. This is the case, for example, of the Palatini and metric-affine formalisms, in which the connection and the metric are regarded as independent fields. These formulations have been explored in the cases of $f(R)$ and Lovelock gravities --- see \eg  \cite{Olmo:2011uz,Sotiriou:2008rp,Exirifard:2007da,Blumenhagen:2012ma}, and it would be natural to extend them to the more general $f($Lovelock$)$ framework.

Another basic aspect omitted in this paper that has been often considered in the cases of $f(R)$ \cite{Dyer:2008hb,Olmo:2011fh,Deruelle:2009pu,Ezawa:2009rh} and Lovelock \cite{Teitelboim:1987zz,Henneaux:1987zz,ChoquetBruhat:1988dw,Deser:2011zk,Ruz:2014ida,MenaMarugan:1992if}\footnote{In fact, this is a subtle topic in the case of Lovelock gravity, the reason being that, in a standard approach, momenta are generically multivalued functions of the time derivatives of the metric, making the Hamiltonian approach ill-defined.} and which could be studied for general $f($Lovelock$)$ theories is the Hamiltonian formulation. 

We also think that it would be interesting to explore how our results on the absence of massive gravitons on m.s.b. for $f($Lovelock$)$ theories extend to less symmetric backgrounds. We already mentioned in the introduction that most of the previous studies on $f($Lovelock$)$ theories were performed in the context of cosmology. It has been in this area that such explorations have been already pursued for certain cosmological backgrounds in the case of $f(R,\mathcal{L}_2)$ theories \cite{DeFelice:2009wp,DeFelice:2010hg}. The theories studied in section \ref{nosca} seem to be particularly relevant in this respect, as they only propagate the usual graviton on m.s.b. It should be possible to clarify whether this property extends to other backgrounds, and what the implications of these results are.

Obviously, constructing additional analytic solutions to these theories and studying their properties would also be a very interesting task, although a challenging one in general. Let us remark that some hope might exist for the class of theories constructed in subsection \ref{nosca}, for which the linearized equations of motion are second-order. In fact, the very same happens for quasi-topological gravity and in that case analytic black hole solutions were built in  \cite{Myers:2010ru} in spite of the higher-derivative and non-topological character of the theory.  A perhaps more doable task which we have somewhat overlooked here would consist in studying the regularity conditions and thermodynamic properties of the $f(\mathcal{L}_n)$ black holes constructed in section \ref{sec:solutions}.

Let us finally mention that, as far as we know, $f($Lovelock$)$ theories have only been considered within the holographic context in \cite{Bueno1,Bueno2} --- rather successfully in that case. It would be interesting to start considering them more often as holographic toy models. This is particularly so for the class of theories constructed in \ref{nosca}. For these, all holographic calculations involving the graviton propagator could be easily performed, given that the only effect of the higher-derivative terms is a change in the normalization of the Newton constant.\footnote{We thank Rob Myers for this remark.} For instance, the coefficient characterizing the stress-tensor two-point function $\ctt$ --- see \eg \cite{Osborn:1993cr} for definitions --- in holographic theories dual to $f($Lovelock$)$ gravities satisfying \req{aaaaa} would be given by 
\begin{equation}
\ctt=\left[1+\lambda\sum_{n=1}^{\lfloor D/2 \rfloor}\partial_n f\left(\bar{\mathcal{L}}\right)\frac{n(D-3)!}{(D-2n-1)!}\,\Lambda^{n-1}\right]\ctt^{ \ssc E}\, ,
\end{equation}
where $\ctt^{ \ssc E}$ is the central charge corresponding to Einstein gravity --- see \eg \cite{Buchel:2009sk}. We leave for future work to further develop the holographic aspects of these theories.

\section*{Acknowledgments}

\noindent We are happy to thank Diego Blas, Patrick Meessen, Rong-Xin Miao, Rob Myers, Gonzalo J. Olmo, and Tom\'as Ort\'in for useful comments and discussions. The work of PB was supported by a postdoctoral fellowship from the Fund for Scientific Research - Flanders (FWO) and partially by the COST Action MP1210 The String Theory Universe. The work of PAC was supported by a ``la Caixa-Severo Ochoa'' International pre-doctoral grant. The work of OLA was supported by the SENESCYT (Secretary of Higher Education, Science, Technology and Innovation of the Republic of Ecuador) scholarship 2014, II. The work of PFR was supported by the Severo Ochoa pre-doctoral grant SVP-2013-067903 and partly by the John Templeton Foundation grant 48222 and the SEV-2012-0249 grant of the ``Centro de Excelencia Severo Ochoa'' program. This work was also supported by the Spanish Ministry of Science and
Education grant FPA2012-35043-C02-01.

\appendix

\section{$f(R,\mathcal{L}_2)$ equations of motion in general dimensions}
\label{frl2}
In this appendix we write down the equations of motion of $f(R,\mathcal{L}_2)$ theories in general dimensions. Observe that this is the most general $f($Lovelock$)$ theory in four dimensions --- the remaining ED identically vanish in that case. We need to compute the quantities appearing in \req{fLL1}. The tensor $\E_{\mu\nu}^{(1)}$ is just the Einstein tensor:
\begin{equation}
\E_{\mu\nu}^{(1)}=R_{\mu\nu}-\frac{1}{2}Rg_{\mu\nu}\, , 
\end{equation}
while
\begin{equation}
P^{(1)\mu\nu}_{\alpha\beta}=-\frac{1}{2}\delta^{\mu\nu}_{\alpha\beta}\, .
\end{equation}
Now, if $D=4$, $\E^{(2)}_{\mu\nu}=0$, while for $D\geq 5$\, , 
\begin{equation}
\E^{(2)}_{\mu\nu}=2RR_{\mu\nu}-4R_{\mu\rho}R_{\nu}^{\rho}+2R_{\alpha\beta\rho\mu}R^{\alpha\beta\rho}_{\ \ \ \ \nu}-4R_{\mu\rho\nu\sigma}R^{\rho\sigma}\, .
\end{equation}
Finally, we have
\begin{equation}
P^{(2)\mu\nu}_{\alpha\beta}=-\delta^{\mu\nu}_{\alpha\beta} R+8\delta^{[\mu}_{[\alpha}R^{\nu]}_{\beta]}-2R^{\mu\nu}_{\alpha\beta}\, ,
\end{equation}
which is also valid in four dimensions. Using this information we can write the equations of motion for $D=4$ and $D\geq 5$ theories respectively as
\begin{equation}
\begin{aligned}
&\frac{\partial f}{\partial R}R_{\mu\nu}-\frac{1}{2}\left[ f-\mathcal{L}_2\frac{\partial f}{\partial \mathcal{L}_2}\right] g_{\mu\nu}+(g_{\mu\nu}\Box-\nabla_{\mu}\nabla_{\nu})\frac{\partial f}{\partial R}-\\
&-4\left[G_{\mu\nu}\Box+\frac{1}{2}R\nabla_{\mu}\nabla_{\nu}-2 R_{\alpha (\mu}\nabla_{\nu)}\nabla^{\alpha}+(g_{\mu\nu}R_{\alpha\beta}+R_{\mu\alpha\beta\nu})\nabla^{\alpha}\nabla^{\beta}\right]\frac{\partial f}{\partial \mathcal{L}_2}=0\, .
\label{eqD4}
\end{aligned}
\end{equation}
\begin{equation}
\begin{aligned}
&\frac{\partial f}{\partial R}R_{\mu\nu}+\frac{\partial f}{\partial \mathcal{L}_2}\Big[2RR_{\mu\nu}-4R_{\mu\rho}R_{\nu}^{\rho}+2R_{\alpha\beta\rho\mu}R^{\alpha\beta\rho}_{\ \ \ \ \nu}-4R_{\mu\rho\nu\sigma}R^{\rho\sigma}\Big]\\
&-\frac{1}{2}\left[ f-\mathcal{L}_2\frac{\partial f}{\partial \mathcal{L}_2}\right] g_{\mu\nu}+(g_{\mu\nu}\Box-\nabla_{\mu}\nabla_{\nu})\frac{\partial f}{\partial R}-\\
&-4\left[G_{\mu\nu}\Box+\frac{1}{2}R\nabla_{\mu}\nabla_{\nu}-2 R_{\alpha (\mu}\nabla_{\nu)}\nabla^{\alpha}+(g_{\mu\nu}R_{\alpha\beta}+R_{\mu\alpha\beta\nu})\nabla^{\alpha}\nabla^{\beta}\right]\frac{\partial f}{\partial \mathcal{L}_2}=0\, .
\label{eqD5}
\end{aligned}
\end{equation}
Observe that these equations reduce to the corresponding $f(R)$ equations of motion when $\partial_2 f=0$. Notice also that a linear term in $\mathcal{L}_2$ gives no contribution in $D=4$ while it does for $D\geq 5$ theories, as expected.
The trace of these equations can be written as
\begin{equation}
\frac{\partial f}{\partial R}R+2\frac{\partial f}{\partial \mathcal{L}_2}\mathcal{L}_2-\frac{D}{2}f+(D-1)\Box\frac{\partial f}{\partial R}+(D-3)(2R\Box-4R_{\mu\nu}\nabla^{\mu}\nabla^{\nu})\frac{\partial f}{\partial \mathcal{L}_2}=0\, ,
\end{equation}
which is a valid expression for $D\geq 4$.

\section{Examples of equivalent scalar-tensor theories}\label{examples}
In this appendix we explicitly compute the equivalent scalar-tensor theories for a couple of classes of $f($Lovelock$)$ theories. The fist example consists of the most general sum of quadratic functions of ED in $D=4$. In the second, we consider a single $f($Lovelock$)$ term consisting of a general product of ED in arbitrary dimensions.

\subsubsection*{Quadratic function}
The most general $f($Lovelock$)$ action containing the usual EH and a negative cosmological constant terms plus quadratic linear combinations of ED in four dimensions reads
\begin{equation}
S=\frac{1}{16\pi G}\int_{\mathcal{M}}d^4x\sqrt{|g|}\, f(R,\mathcal{L}_2)\, , 
\end{equation}
where
\begin{equation}
f(R,\mathcal{L}_2)=\frac{6}{\tilde{L}^2}+R+\alpha \tilde{L}^2R^2+\beta \tilde{L}^4 R\mathcal{L}_2+\gamma \tilde{L}^6 \mathcal{L}_2^2\, ,
\end{equation}
and where $\alpha$, $\beta$ and $\gamma$ are dimensionless constants. The Hessian matrix of $f(R,\mathcal{L}_2)$ reads in this case:
\begin{equation}
H(f)=
\begin{pmatrix}
2\alpha \tilde{L}^2 & \beta \tilde{L}^4\\
\beta \tilde{L}^4 & 2\gamma \tilde{L}^6\\
\end{pmatrix}\, .
\end{equation}
Leaving the trivial case $\alpha=\beta=\gamma=0$ aside, we see that:
\begin{equation}
\operatorname{rank}(H)=
\begin{cases}
2 & \mbox{if } 4\alpha \gamma-\beta^2\neq 0\, ,\\
1 & \mbox{if } 4\alpha \gamma-\beta^2=0\, .
\end{cases}
\end{equation}
Hence, according to our analysis in the main text, in the first case we need to introduce two scalars, while in the second it is enough with a single one. We have the function 
\begin{equation}
f(\phi_1,\phi_2)=\frac{6}{\tilde{L}^2}+\phi_1+\alpha \tilde{L}^2\phi_1^2+\beta \tilde{L}^4 \phi_1\phi_2+\gamma \tilde{L}^6\phi_2^2\, .
\end{equation}
Then we define
\begin{eqnarray}
\varphi_1&=&\frac{\partial f}{\partial \phi_1}=1+2\alpha \tilde{L}^2\phi_1+\beta \tilde{L}^4 \phi_2\, ,\\
\varphi_2&=&\frac{\partial f}{\partial \phi_2}=\beta \tilde{L}^4 \phi_1+2\gamma \tilde{L}^6 \phi_2\, .
\end{eqnarray}
If $4\alpha \gamma-\beta^2\neq 0$ then these fields are independent. The Legendre transform of $f$ reads
\begin{equation}
\tilde V(\varphi_1,\varphi_2)=-\frac{6}{\tilde{L}^2}+\frac{1}{\tilde{L}^6(4\alpha \gamma-\beta^2)}\left(\gamma \tilde{L}^4 (\varphi_1-1)^2-\beta \tilde{L}^2 (\varphi_1-1)\varphi_2+\alpha \varphi_2^2\right)\, .
\end{equation}
Therefore, the equivalent scalar-tensor theory is
\begin{equation}
S'=\frac{1}{16\pi G}\int_{\mathcal{M}}d^4x\sqrt{|g|}\left[\frac{6}{\tilde{L}^2}+\varphi_1 R+\varphi_2 \mathcal{L}_2-\frac{\gamma \tilde{L}^4 (\varphi_1-1)^2-\beta \tilde{L}^2 (\varphi_1-1)\varphi_2+\alpha \varphi_2^2}{\tilde{L}^6(4\alpha \gamma-\beta^2)}\right]\, .
\end{equation}
Of course, this does not work if $4\alpha\gamma-\beta^2=0$. 
In that case, the quadratic term is a perfect square of the form
\begin{equation}
S=\frac{1}{16\pi G}\int_{\mathcal{M}}d^4x\sqrt{|g|}\left[\frac{6}{\tilde{L}^2}+R+\lambda \tilde{L}^2(R+c\tilde{L}^2 \mathcal{L}_2)^2\right]\, ,
\end{equation}
where $c$ is some unimportant constant. We find that this is equivalent to the following scalar-tensor theory with one single scalar $\varphi$
\begin{equation}
S'=\frac{1}{16\pi G}\int_{\mathcal{M}}d^4x\sqrt{|g|}\left[\frac{6}{\tilde{L}^2}+\varphi R+\varphi c \tilde{L}^2\mathcal{L}_2-\frac{(\varphi-1)^2}{4\lambda \tilde{L}^2}\right]\, .
\end{equation}

\subsubsection*{General product of ED}
Let us consider now the following action:
\begin{equation}
S=\frac{1}{16\pi G}\int_{\mathcal{M}}d^Dx\sqrt{|g|}\left[-2\Lambda_0 +R+\lambda\prod_{i=1}^n\mathcal{L}_{p_i}^{v_i}\right],
\end{equation}
where $\{p_i\}_{i=1}^n\subset \{1,2,\dots,\lfloor D/2 \rfloor\}$ and $v_i\in \mathbb{Z}-\{0\}$ are non-zero exponents. This action contains a rather generic $f($Lovelock$)$ term, namely, one consisting of a product of ED. Let us also assume that $p_1=1$, so there is a power of $R$ in the product. The Hessian matrix of $f(\mathcal{L})=-2\Lambda_0 +R+\lambda\prod_{i=1}^n\mathcal{L}_{p_i}^{v_i}$ reads
\begin{equation}
H_{ij}=(v_iv_j-v_i\delta_{ij})\frac{\lambda\prod_{i=1}^n\mathcal{L}_{p_i}^{v_i}}{\mathcal{L}_{p_i}\mathcal{L}_{p_j}}\, ,
\end{equation}
whose rank can be seen to be given by

\begin{equation}
r=
\begin{cases}
n & \mbox{if }\sum_{i=1}^n v_i \neq 1,\\
n-1 & \mbox{if } \sum_{i=1}^n v_i =1.
\end{cases}
\end{equation}
The second case can happen if we allow the exponents to be non-integer or if some of them are negative. In the first case we can compute the Legendre transform of the function $f(\chi_1,\dots,\chi_n)=-2\Lambda_0+\chi_1+\lambda\prod_{i=1}^n\chi_i^{v_i}$. The transformed fields are
\begin{equation}
\phi_1=\partial_1 f(\chi)=1+\frac{v_1\lambda}{\chi_1} \prod_{j=1}^n\mathcal{L}_{p_j}^{v_j},\quad \phi_i=\partial_i f(\chi)=\frac{v_i\lambda}{\chi_i}\prod_{j=1}^n\mathcal{L}_{p_j}^{v_j},\quad i>1,
\end{equation}
so we find:
\begin{equation}
\tilde V(\phi)=2\Lambda_0+(s-1)\lambda^{\frac{1}{1-s}}\left(\frac{\phi_1-1}{v_1}\right)^{\frac{v_1}{s-1}}\prod_{j=2}^n\left(\frac{\phi_j}{v_j}\right)^{\frac{v_j}{s-1}},
\end{equation}
where $ s= \sum_{i=1}^n v_i$. Therefore, this theory is equivalent to
\begin{equation}
S'=\frac{1}{16\pi G}\int_{\mathcal{M}}d^Dx\sqrt{|g|}\left[-2\Lambda_0+\sum_{i=1}^n\phi_i \mathcal{L}_{p_i}-(s-1)\lambda^{\frac{1}{1-s}}\left(\frac{\phi_1-1}{v_1}\right)^{\frac{v_1}{s-1}}\prod_{j=2}^n\left(\frac{\phi_j}{v_j}\right)^{\frac{v_j}{s-1}}\right].
\end{equation}
If $s=1$, the Legendre transform is constant $\tilde V=2\Lambda_0$, and we have the constraint
\begin{equation}
\left(\frac{\phi_1-1}{v_1}\right)^{v_1}\prod_{i=2}^n\left(\frac{\phi_i}{v_i}\right)^{v_i}=\lambda,
\end{equation}
from which we can extract, for example, $\phi_n$ as a function of the rest of the fields:
\begin{equation}
\phi_n=(1-s')\lambda^{\frac{1}{1-s'}}\left(\frac{\phi_1-1}{v_1}\right)^{\frac{v_1}{s'-1}}\prod_{j=2}^{n-1}\left(\frac{\phi_j}{v_j}\right)^{\frac{v_j}{s'-1}},
\end{equation}
where now, $ s'= \sum_{i=1}^{n-1} v_i$. Hence, in the case in which $s=1$ --- which means that $f$ is a homogeneous function of degree 1 --- the theory is equivalent to the following scalar-Lovelock theory with $n-1$ scalar fields and without scalar potential 
\begin{equation}
S'=\frac{1}{16\pi G}\int_{\mathcal{M}}d^Dx\sqrt{|g|}\left[-2\Lambda_0+\sum_{i=1}^{n-1}\phi_i \mathcal{L}_{p_i}+(1-s')\lambda^{\frac{1}{1-s'}}\left(\frac{\phi_1-1}{v_1}\right)^{\frac{v_1}{s'-1}}\prod_{j=2}^{n-1}\left(\frac{\phi_j}{v_j}\right)^{\frac{v_j}{s'-1}}\mathcal{L}_{p_n}\right].
\end{equation}

\renewcommand{\leftmark}{\MakeUppercase{Bibliography}}
\phantomsection
\bibliographystyle{JHEP}
\bibliography{fLovelockPaper}
\label{biblio}

\vspace{1cm}
\noindent\rule{4cm}{0.4pt}\\\\
E-mail:\\
\url{pablo@itf.fys.kuleuven.be} \\\url{ pablo.cano@uam.es}  \\\url{oscar.lasso@estudiante.uam.es} \\\url{p.f.ramirez@csic.es}

\end{document}